\documentclass[aps,prl,twocolumn,superscriptaddress,floatfix,letter,longbibliography]{revtex4-2}

\usepackage{epsfig,amsmath,amssymb,color,amsfonts,physics,microtype}
\usepackage[bookmarks=true,colorlinks,citecolor=red]{hyperref}
\usepackage{dsfont}
\usepackage{wrapfig}
\usepackage{tikz}
\usepackage{physics}
\usepackage{mathrsfs}
\usepackage[export]{adjustbox}
\usetikzlibrary{quantikz}
\usepackage{soul}
\usepackage{comment}
\usepackage{ulem}

\definecolor{darkgreen}{RGB}{1,151,0}

\begin{document}

\title{Finite-temperature formation of magnetic plateaus and simplex liquid states on the frustrated ruby lattice}

\author{Antonio Francesco Mello}
\affiliation{Center for Computational Quantum Physics, Flatiron Institute, New York, NY, 10010, USA}
\affiliation{International School for Advanced Studies (SISSA), via Bonomea 265, 34136 Trieste, Italy}
\author{E.\ Miles Stoudenmire}
\author{Joseph Tindall}
\affiliation{Center for Computational Quantum Physics, Flatiron Institute, New York, NY, 10010, USA}

\begin{abstract}
Geometric frustration in quantum systems can stabilize unconventional phases of matter that avoid traditional magnetic ordering at low temperatures.
Here, we observe this phenomenon while mapping out the finite temperature phase diagram of the spin-1/2 Heisenberg antiferromagnet on the ruby lattice with next-nearest-neighbor interactions. Using an infinite tensor network state (iTNS) optimized and measured with belief propagation (BP) and corrections to BP, we observe the low temperature formation of stable magnetic plateaus at various magnetic field strengths. We find these plateaus host a novel `simplex liquid state' -- a disordered phase involving strongly paired spin simplices that retains non-zero residual entropy due to an exponentially large subspace of crystalline configurations. 
We accurately quantify the energy gap associated with these states and show that, as the temperature of the system is lowered, it does not go through a phase transition to reach them: the heat capacity remains finite and continuous at all observed temperatures. 
Our work demonstrates how BP-based tensor network techniques provide a powerful route to understanding frustrated quantum magnets at finite temperature.
\end{abstract}

\maketitle

\paragraph*{Introduction. ---}
Geometric frustration has long been recognized as a key mechanism for stabilizing unconventional phases in quantum magnets, including quantum spin liquids~\cite{wen2004quantum, savary2017quantum} and highly correlated states~\cite{read1991frustr, sachdev1992antif, anderson1987resonating, balents1998nodal}. Understanding how such states form at finite temperatures and avoid conventional magnetic ordering remains an outstanding challenge.

The ruby lattice ~\cite{ghosh2025simplexcrystalgroundstate, maity2024gappedgaplessquantumspin,schmoll2024bathing, farnell2011spin,schafer2023abundance}, a highly frustrated two-dimensional network that mimics the core structure of corner-sharing geometries such as the three-dimensional pyrochlore lattice ~\cite{moessner98spinliquid,gardner2010magnetic,iqbal2019quantum}, presents a natural environment for understanding finite temperature frustrated systems. This geometry has recently been implemented in programmable simulators based on Rydberg atoms, providing a platform for the realization of topologically ordered quantum spin liquids (QSL)~\cite{semeghini2021probing, giudici2022dynamical, tarabunga2022gauge, verresen2021prediction}. Moreover, a slightly distorted version of the lattice is realized in layered transition-metal fluorides such as ${\rm Cs}{\rm Ba}{\rm Fe}_{3}{\rm F}_{12}$~\cite{jeschke2026}.

The zero-temperature ($T$ = 0) ground state of the next-nearest-neighbor Heisenberg model on the ruby lattice was recently shown to exhibit fractionalized magnetic plateaus~\cite{ghosh2025simplexcrystalgroundstate}, using well-established tensor network techniques ~\cite{jiang2008accurate, nishino1996corner, orus2009simulation, Yang_2026}. This initial investigation suggested the plateaus host discrete crystalline phases that break the sixfold rotational symmetry of the lattice. However, these $T = 0$ insights left many open questions such as the exact ground state degeneracy, the finite-temperature stability of these plateaus, and their associated thermodynamic signatures. Performing simulations at finite temperature is necessary to resolving these unknowns and, more broadly, connecting computational simulations with experimental realizations where thermal and quantum fluctuations compete.

\begin{figure*}
\centering
\includegraphics[width=\textwidth]{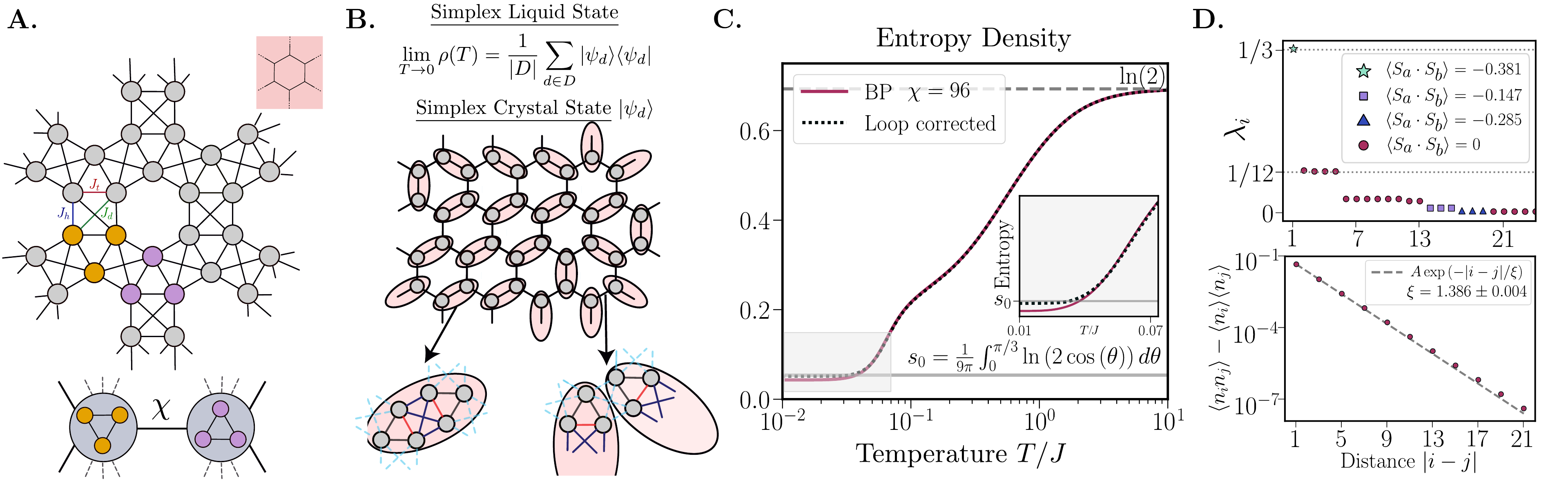}
\caption{\textbf{A)} Section of the ruby lattice with nearest ($J_t, J_h$) and next-to-nearest neighbor ($J_d$) Heisenberg interactions---see Eq.~(\ref{eq:hamiltonian}). Here, we study the isotropic $J_{d} = J_{t} = J_{h} = 1$ case. Upon grouping of the vertices into triangles (simplices), the lattice becomes a honeycomb with coordination number $z=3$. Below: Two-site iTNS (infinite tensor network state) used in our calculations to represent the finite temperature density matrix in the thermodynamic limit. Each tensor possesses three virtual indices of dimension $\chi$ and six physical indices. \textbf{B)} Illustration of the liquid state formed at low temperatures involving a perfect mixture of crystalline states with the strongly paired (dark blue bonds) spin simplices forming a dimer covering of the hexagonal lattice and weaker correlations (dashed light blue bonds) manifest between other pairs of simplices. \textbf{C)} Entropy per spin as a function of $T/J$. In the $T\to 0$ limit, the loop corrected residual entropy is in good agreement with that for dimer coverings of the honeycomb lattice \cite{Kastelyn1961, Temperley1961}. \textbf{D)} Upper: eigenvalues of the reduced density matrix for a pair of simplices at $T/J=0.005$. The leading $\lambda_{1} = 1/3$ value directly corresponds to a state $\vert \psi_{1} \rangle$ in which the two simplices are strongly paired together. The eigenvalues are colored by the value of the correlator $\langle \mathbf{S}_a \cdot \mathbf{S}_b \rangle$ between interacting spins $a$ and $b$ in the two different triangles of $\vert \psi_{i} \rangle$. Lower: connected correlator for the occupation operator $n_{i} = \left(\vert \psi_{1} \rangle \langle \psi_{1} \vert \right)_{i}$ showing an exponential decay with the distance.}\label{Fig:h0_panels}
\end{figure*}

In this work, by capitalizing on recent developments in tensor networks based on belief propagation---that make two-dimensional calculations more affordable and scalable while offering controlled corrections---we accurately address the thermodynamics of a frustrated two-dimensional magnet over temperatures spanning many orders of magnitude.
Specifically, we use a purified infinite tensor network state (iTNS) and leverage the belief propagation (BP)~\cite{alkabetz2021tensor, tindall2023gauging} contraction algorithm and corrections thereof \cite{evenbly2026loop} to realize the finite temperature states of the next-nearest-neighbor Heisenberg model on the ruby lattice down to very low temperatures, directly in the thermodynamic limit. Upon cooling the system, we observe the anticipated formation of the $m_z = 0, 1/3, 1/2$ and $2/3$ plateaus. 
Crucially, our methodology allows us to quantify the thermal extent and energy gap associated with these plateaus.

Surprisingly, we find the system does not undergo a phase transition to form these stable magnetic phases, with the heat capacity remaining smooth and continuous at all temperatures. Instead, the residual entropy remaining in the system down to very low temperatures is observed to be numerically consistent with an exponentially degenerate manifold of crystalline states $\vert \psi_{d} \rangle$ involving regularly spaced, correlated simplex pairs. The resulting mixed state we observe is a stable `simplex liquid', or a perfect disordered mixture of all of these states $\rho \propto \sum_{d \in D}\vert \psi_{d} \rangle \langle \psi_{d} \vert$. 

Although the core of these magnetic plateaus forms smoothly without any conventional phase transition, at finite temperatures on the edge of the $m_{z} = 1/2$ plateau, we observe thermally-driven quantum fluctuations which destabilize the $m_{z} = 1/2$ liquid phase and cause the system to discontinuously switch into a different liquid state. This switching is heralded by a `lambda' peak in the heat capacity, and the associated quantum fluctuations appear to stem from a nearby zero-temperature quantum phase transition.

Our numerical methodology, which builds upon recent, significant advances within the PEPS/TNS community~\cite{alkabetz2021tensor, tindall2023gauging, evenbly2026loop, evenbly2025partitioned, guo2023block, midha2025beyond, midha2026belief, tindall2026}, is key to our findings. 
Specifically, by using efficient contraction schemes such as BP, a first order cluster correction,  and GPU hardware,  we are able to reach the bond dimensions ($\chi \sim 200$) in the iTNS required to accurately measure thermodynamic observables upon cooling.
This approach here opens up a broad pathway to studying finite temperature frustrated two- and three-dimensional systems with tensor networks.

\paragraph*{Model. ---}
We consider the antiferromagnetic Heisenberg Hamiltonian with an external magnetic field $h$
\begin{equation}\label{eq:hamiltonian}
 H = \sum_{k \in \{t,d,h\}}\sum_{\langle i,j\rangle_k} J_k \textbf{S}_i \cdot \textbf{S}_j - h \sum_i S^z_i,
\end{equation}
where $\textbf{S}_i = (S^{x}_{i}, S^{y}_{i}, S^{z}_{i})^{T}$ and $S^{\alpha}_{i} = \frac{1}{2}\sigma^{\alpha}_{i}$. The subscript $k=t,d,h$ denotes the nearest-neighbor couplings $J_t$, $J_h$, and the next-nearest-neighbor $J_d$. See Fig.~\ref{Fig:h0_panels}A for a visual representation of the model. In the following, we will focus on the isotropic case $J_t = J_h = J_d = J$ and leave the anisotropic case which spans setups such as the maple-leaf or nearest-neighbor ruby lattice \cite{farnell2011spin, Sch_fer_2026, ebert2026} to future work.
A previous study~\cite{ghosh2025simplexcrystalgroundstate} analyzed the rich ground-state physics of Eq. (\ref{eq:hamiltonian}). At zero field $h/J=0$, three non-magnetic `simplex crystal' ground states were observed involving regularly spaced, strongly correlated pairs of spin triangles (simplices). Specifically, the regular arrangement ensures that each triangle of spins forms a strongly correlated low energy state with one of the three neighboring triangles whilst weak correlations with the other two neighbors allow the system to lower its energy below that of a simple product state of simplex pairs. 

Upon increasing the magnetic field, Ref.~\cite{ghosh2025simplexcrystalgroundstate} further showed that the magnetic response of the ground state exhibits plateaus at values $m_z = 0,1/3, 1/2, 2/3$ where $m_z =  \lim_{n \rightarrow \infty}\frac{1}{n}\sum_{i=1}^{n}\langle \sigma^{z}_{i} \rangle$ is the magnetization density. It was argued that these plateaus host magnetized simplex crystal states that, in addition to breaking the $C_6$ symmetry, also break the reflection one with respect to the $\sigma_d$ planes.

As we will demonstrate in our finite temperature results, within these plateaus the low energy subspace of the system actually contains an exponentially large number of these crystalline states. We find the system forms a perfect disordered liquid from these states, with no breaking of any lattice symmetries even down to temperatures as low as $T/J \sim 10^{-3}$. The finite degeneracy and crystalline nature of the ground states observed in Ref.~\cite{ghosh2025simplexcrystalgroundstate}  likely stemmed from the size of the unit cell used (one hexagonal ring) and the translationally-invariance broken initial states used to seed the algorithm.


\paragraph*{Methods. ---}
To address the finite temperature properties of $H$, we use an iTNS ansatz to represent the \textit{unnormalized} square root of the equilibrium state $\rho^{\frac{1}{2}}(\beta) = \sqrt{\exp(- \beta H)} = \exp(- \beta H /2)$ with $\beta = \frac{1}{T}$. We first group the vertices of the triangles of the ruby lattice and work with a coarse-grained honeycomb geometry. Our iTNS is then a two-site unit cell as shown in Fig.~\ref{Fig:h0_panels}A, with each tensor having six external indices (appropriate for a mixed state of three spins) and three virtual indices.

The iTNS is initialized in the infinite temperature state $\rho^{\frac{1}{2}}(0) = \mathcal{I}$ which has bond dimension $\chi = 1$ and finite temperature is reached by cooling the system in imaginary time $\rho^{\frac{1}{2}}(\beta) = \left(\prod_{j=1}^{N}\exp(- \frac{\delta \beta}{2}H)\right)\rho^{\frac{1}{2}}(0)$, using a second order Trotterization of the propagator $\exp(- \frac{\delta \beta}{2} H)$ with $\delta \beta \cdot N = \beta $. We set $\delta \beta \cdot J = 10^{-3}$ throughout to minimize Trotterization errors. 
More details on gate applications and state truncation are provided in the Appendix and Ref.~\onlinecite{tindall2023gauging}.

 Our approach allows us to measure observables via two methods. The first of these is via the standard formula $\displaystyle \langle O \rangle = {\rm Tr}\left(\rho(\beta) O\right) /{\rm Tr}\left(\rho(\beta)\right)$, which can be calculated under the BP approximation by running BP on the norm network $Z= {\rm Tr}\left(\rho(\beta)\right)$ to obtain converged BP message tensors, inserting those incident to the region of support of $O$ and then computing the corresponding ratio (see Appendix). The second is to recognize that the Bethe free energy of $Z$, which can be evaluated by local contraction of the BP messages tensors and the local tensors in the iTNS, is the BP approximation to the free energy density $f(\beta)= - \lim_{n \rightarrow \infty}\frac{1}{n} {\rm ln} \left( {\rm Tr}(\exp(- \beta H )) \right) $. Taking appropriate derivatives of this quantity then gives access to the relevant thermodynamic observables \footnote{All derivatives in this work are computed using a second-order finite-difference scheme.}. 

The free energy approach is particularly useful as it allows one to leverage recent developments in the contraction of tensor networks~\cite{evenbly2025partitioned, evenbly2026loop} and make a loop-cluster correction to improve the accuracy of the results beyond the BP approximation, accounting for correlations around the hexagons of the lattice (see Appendix).

\begin{figure}[t!]
\centering
\includegraphics[width=\columnwidth]{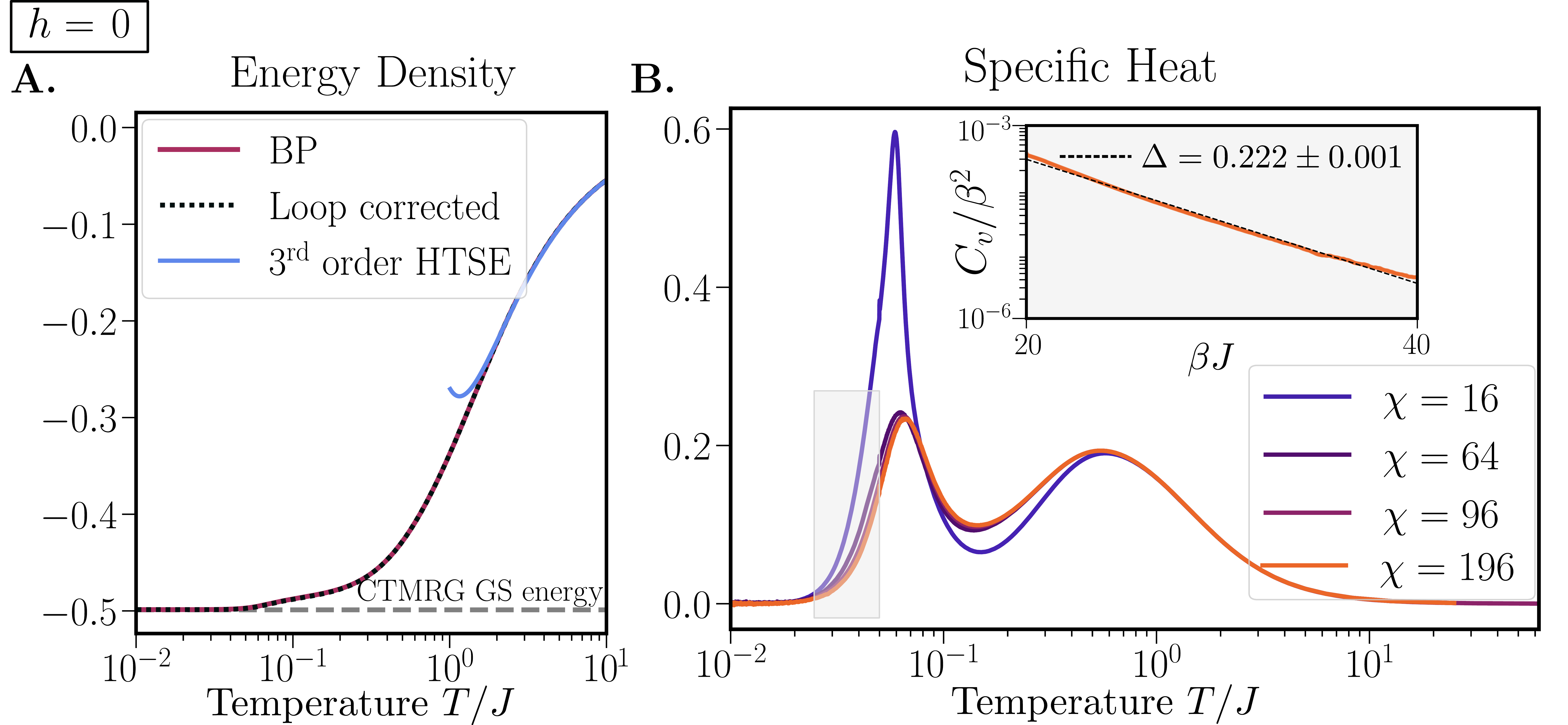}
\caption{\textbf{A)} Energy density $e/J$ as a function of temperature measured with both BP and a first order loop correction with $\chi = 96$. At low temperatures, they match the Ref.~\cite{ghosh2025simplexcrystalgroundstate} for the ground state, whilst the high temperature behavior matches a third-order high temperature series expansion (HTSE) --- see Eq.(\ref{Eq:HTSE}). \textbf{B)} Specific heat $C_v$ as a function of temperature for different values of the bond dimension $\chi$. Inset: estimate of the gap $\Delta \approx 0.22J$ via the low temperature  fit $C_{v} \beta^{2} = Ae^{- \Delta \beta}$.}
\label{Fig:h0_energy_cv}
\end{figure}

\begin{figure*}
\centering
\includegraphics[width=\textwidth]{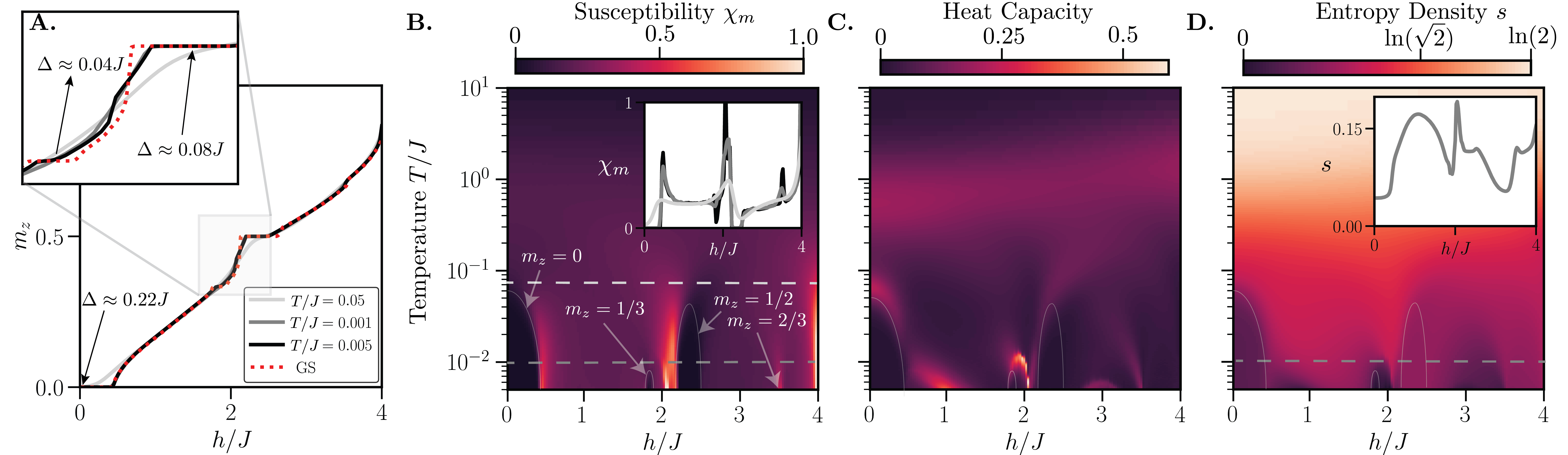}
\caption{\textbf{A)} Magnetization profile as a function of the external field $h$ at different temperatures $T/J$. Finite temperature and ground state results are obtained through $\chi = 96$ and $\chi =12$ iTNS, respectively. \textbf{B)} Heatmap for the magnetic susceptibility $\chi_m$. The dark regions (whose boundaries are marked) are associated with the formation of the magnetic plateaus and simplex liquid states. \textbf{C)} Heatmap for the specific heat $C_v$. \textbf{D)} Heatmap for the entropy density. Inset: entropy density as a function of the magnetic field $h$ for $T/J=0.005$.}
\label{Fig:HeatMaps}
\end{figure*}

\paragraph*{Results, zero magnetic field ---}

We begin by investigating the thermodynamic properties of the system in the absence of an external field, i.e. $h = 0$. We cool the system from infinite temperature down to $T/J=10^{-2}$, measuring various thermodynamic quantities including the energy density $e = \frac{\partial f}{\partial \beta} = \lim_{n \rightarrow \infty}\frac{1}{n}\langle H \rangle$, the specific heat $C_{v} (\beta) = - \beta^{2} \frac{\partial e (\beta)}{\partial \beta}$, and the entropy density $s(\beta) = s(0) - \int_{0}^{\beta} \left[e(\beta') - e(\beta) \right] d \beta'$
    as a function of temperature, where $s(0)= \ln(2)$ for a system of $S=1/2$ spins. We compute both the BP approximation to these quantities and the first-order cluster correction via the corrected free energy. 
    
In Fig.~\ref{Fig:h0_panels}C we show how the residual entropy density $s$ saturates at low temperature to a non-zero value $s_{0}$. We find, upon using a cluster correction to account for the local hexagonal structure of the lattice, that $s_{0} \approx \frac{1}{3}\lambda$, where $\lambda = \frac{1}{3 \pi}\int_{0}^{\frac{\pi}{3}}\ln(2 \cos(\theta))d\theta$ \cite{Kastelyn1961, Temperley1961} is the residual entropy of dimer coverings of the honeycomb lattice. Moreover, we observe $\langle \mathbf{S}_{i} \cdot \mathbf{S}_{j} \rangle$ takes an identical value on every inter-simplex bond at all temperatures and so there is no local order parameter that can identify the simplex pairing in the system.

These findings indicate that at low temperatures, the system has effectively formed a disordered `simplex liquid state', i.e.\ a  translationally-invariant classical mixture of an exponential number of crystalline states $\vert \psi_{d} \rangle$ such as that illustrated in Fig. \ref{Fig:h0_panels}B.  
The number of these crystalline  states is equal to the number of dimer coverings of the hexagonal lattice: the fact $\lambda / s_{0} = 3$ is because we are measuring an entropy per spin and each triangle / honeycomb vertex in our model contains three spins.
The energy density of $e/J = -0.498..$ (see Fig. \ref{Fig:h0_energy_cv}A) we measure for this liquid state agrees, to three decimal places, with the energy density of the three specific crystalline configurations the $T = 0$ calculation of Ref.~\cite{ghosh2025simplexcrystalgroundstate} identified.
    
We emphasize that our results are completely unchanged upon using a larger, six-site ($18$-spin) unit cell, where the system would be capable of forming one of the three specific crystalline states observed in Ref.~\cite{ghosh2025simplexcrystalgroundstate} if it were energetically favorable over the liquid state. Thus, we conclude that the system preferentially forms the liquid state in the absence of any perturbations to the Hamiltonian $H$ in Eq. (\ref{eq:hamiltonian}).

We gain deeper insights into the nature of the state from an analysis of the spectrum of the two-site (six spin) BP-computed reduced density matrix \newline $\rho_{v_1, v_2}(\beta) = {\rm Tr}_{v \neq (v_1, v_2)}(\rho (\beta))$
at low temperatures across one of the edges $e = v_1 \leftrightarrow v_2$ of the hexagonal lattice. We plot the eigenvalues in the upper panel of Fig.~\ref{Fig:h0_panels}D. We find a leading eigenvalue $\lambda_{1}$ which corresponds to a correlated six-spin state $\vert \psi_{1} \rangle$ with an energy density of $e_{i}/J = -0.485...$. The fact that $\lambda_1 \approx 1/3$ reflects how the edge is expected to be covered in $1/3$ of all dimer coverings of the hexagonal lattice and strongly corroborates our picture of the disordered simplex liquid state in Fig.~\ref{Fig:h0_panels}B. We find the state $\vert \psi_{1} \rangle$ is numerically very close to the ground state of the Hamiltonian in Eq.(\ref{eq:hamiltonian}) for a single pair of spin simplices.
    
The other eigenstates have eigenvalues adding up to $2/3$ and correspond to the case of the two triangles being in a reduced state orthogonal to the paired one and thus choose to pair more strongly with different simplices instead. The first four such states have degenerate eigenvalues $\lambda_2 = 1/12$ and have the property $\langle \mathbf{S}_{a} \cdot \mathbf{S}_{b} \rangle = 0$ for spins $a$ and $b$ which are not in the same triangle. There are a handful of density matrix eigenstates, however, where the triangles remain weakly correlated $\langle \mathbf{S}_{a} \cdot \mathbf{S}_{b} \rangle < 0$ and these states are key in allowing the system to lower its energy density below $e/J = -0.485...$ (we find $e/J = -0.498...$ at $T/J = 10^{-2}$) where the slightly higher energy would be expected if the $\vert \psi_{d} \rangle$ were just product states of the low energy simplex pairs $\vert \psi_{1} \rangle$. The presence of these weakly correlated states corroborates the intuition of Ref.~\cite{ghosh2025simplexcrystalgroundstate} that weak quantum correlations between unpaired simplices help a given crystalline state $\vert \psi_{d} \rangle$ to lower its energy.
    
In the lower panel of Fig.~\ref{Fig:h0_panels}, we show the behavior of the connected correlator associated with the simplex pair occupation operator $n_i = (\vert \psi_{1} \rangle \langle \psi_{1} \vert)_{i}$ as a function of the number of hexagonal bonds $\vert i - j \vert$ between bonds $i$ and $j$ along a path of the hexagonal lattice. Fitting an exponential decay, we obtain a correlation length $\xi \approx 1.4$, which is significantly less than the size of a hexagon and explains the accuracy of the BP approximation in our calculations.
    
Figure~\ref{Fig:h0_energy_cv}A shows the energy density as a function of temperature. We find its high-temperature ($T/J < 1$) behavior is in good agreement with a third order high-temperature series expansion  \cite{pierre2024high}
\begin{equation}
        e = -\frac{9}{16}\beta + \frac{3}{16}\beta^{2} + \frac{27}{256}\beta^{3} + \mathcal{O}(\beta^{4}).
        \label{Eq:HTSE}
\end{equation} 
and the lower temperature $T/J = 10^{-2}$ value is consistent with the corner transfer matrix renormalization group (CTMRG)-based $T = 0$ calculation of Ref. \cite{ghosh2025simplexcrystalgroundstate} to three decimal places as discussed above.
 
The specific heat $C_v$ is displayed in Fig.~\ref{Fig:h0_energy_cv}B for iTNS bond dimensions up to $\chi = 196$, along with a numerical estimate for the gap $\Delta = 0.22(1)J$ of the state obtained by fitting the low temperature behavior with the anticipated form $C_{v} \sim \beta^2 \exp(-\beta \Delta)$ for a gapped system. We find the heat capacity and estimate for the gap to converge around a bond dimension of $\chi \sim 100$. Most strikingly, the specific heat curve stays continuous and finite at all temperatures and the low temperature peak is not diverging with increased bond dimension---one would expect the latter if the system were to go through a finite-temperature phase transition upon cooling down to the plateau. This result ties in with the understanding that the system reaches a disordered liquid state with non-zero residual entropy and which does not break the discrete symmetries of the hexagonal lattice.

\begin{figure}
\centering
\includegraphics[width=\columnwidth]{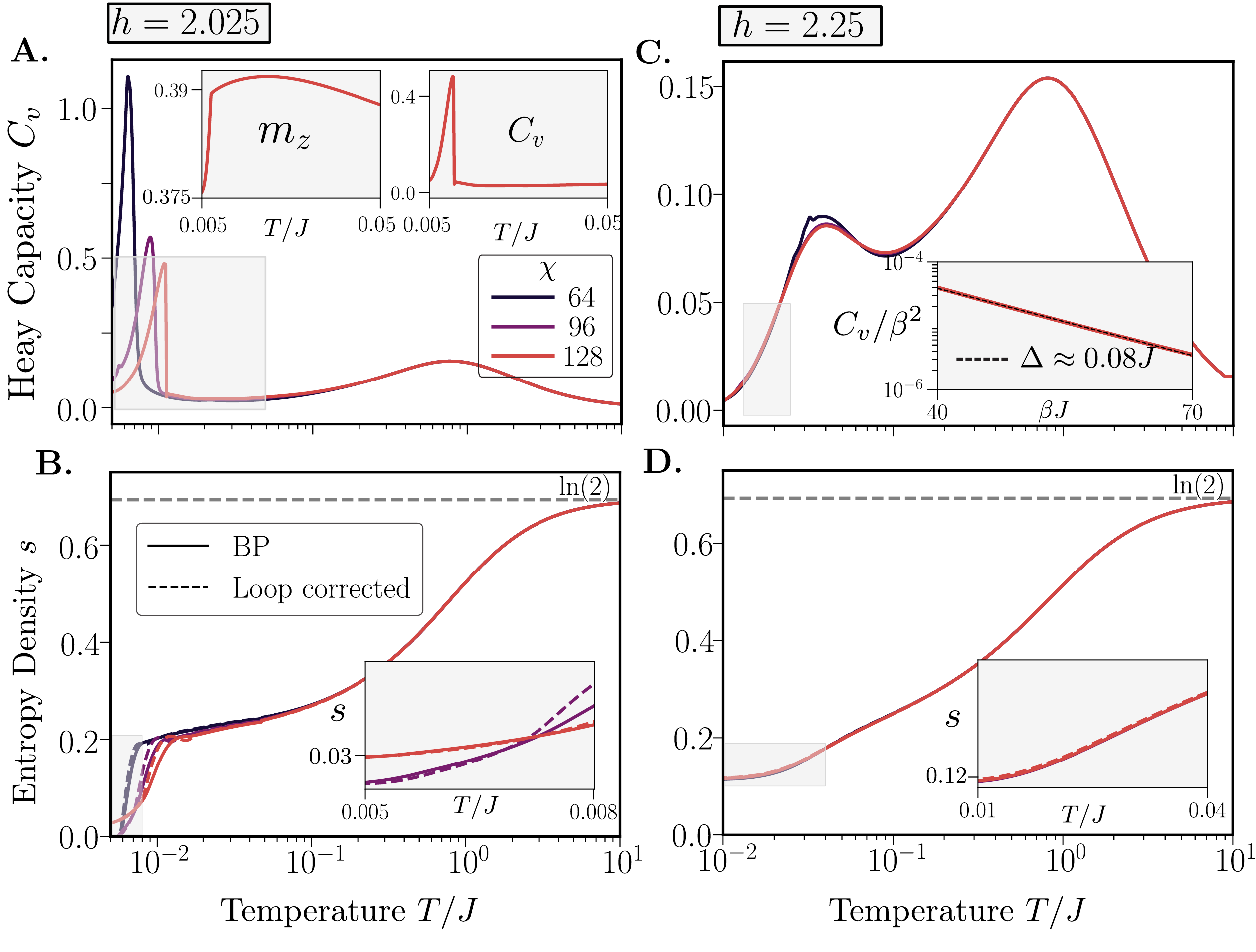}
\caption{\textbf{A)} Specific heat versus temperature at the interface between the $m_z = 1/3$ and $m_z = 1/2$ plateaus for various bond dimensions. Insets: magnetization density and specific heat on a linear temperature scale near the switching point. \textbf{B)} Entropy versus temperature and low temperature inset.  \textbf{C)} Specific heat versus temperature for a field strength within the $m_z=1/2$ plateau, displaying analogous features to the $h=0$ case. Inset shows extraction of the energy gap via fitting $C_{v} \propto \beta^{2} e^{-\Delta \beta}$ in the low temperature limit.
\textbf{D)} Entropy as a function of temperature. The residual entropy $s(T)$ as $T \rightarrow 0$, which is converged in bond dimension and weakly increased by a loop correction, signals a massive extensive degeneracy of the low energy manifold.}
\label{Fig:SpecificMags}
\end{figure}

\paragraph*{Results, Finite Magnetic Field ---}
We now investigate the effect of a magnetic field in the $z$ direction on the behaviour of the system.
We measure the local magnetization density $m_{z} $, the
magnetic susceptibility $\displaystyle \chi_m  = \frac{\partial m_z}{\partial h}$, the specific heat and the entropy as a function of both temperature and the field strength. Figure ~\ref{Fig:HeatMaps} displays our full results and Fig. \ref{Fig:SpecificMags} shows entropies and heat capacity curves for two different field strengths.  The results indicate a broad range of simplex liquid states with different magnetic orderings are manifest in the system at low temperature and more fragile liquid states are likely to emerge at even lower temperatures.

Firstly, for small fields $h$ and low temperatures we find the non-magnetic simplex liquid identified earlier remains stable, with a consistent residual entropy, vanishing magnetic susceptibility
and non-diverging double peak structure in the heat capacity. Beyond $h_{c} \sim 0.4J$, a low temperature response to the field develops with a sequence of low temperature magnetic plateaus $m_{z} = 1/3, 1/2, 2/3$ emerging at increasing field strengths consistent with Ref. \cite{ghosh2025simplexcrystalgroundstate} and an independent ground-state calculation we performed (red dotted line, Fig.~\ref{Fig:HeatMaps}(a)) via imaginary time evolution on a pure initial state.

The $m_z = 0$ plateau forms at the highest temperatures among the plateau states, at around $T \sim 0.06\, J$.
The next highest energy scale corresponds to the $m_{z} = 1/2$ plateau, spanning $2.2 \lesssim h/J \lesssim 2.525$, and it is clearly resolved from the susceptibility and heat capacities. The $m_{z} = 1/2$ plateau has a very large residual entropy, around double that of the $m_z = 0$ plateau. Consistent with a large remaining entropy, we observe that a very shallow heat capacity peak heralds the formation of this plateau.
Again, cluster corrections only become significant in non-local quantities such as the entropy at very low temperatures, where they slightly raise the residual entropy above the BP prediction.

The $m_{z} = 1/3$ plateau, meanwhile, is much less stable to thermal fluctuations and appears only in a narrow field range $1.8 \lesssim h/J \lesssim 1.875$, requiring temperatures $T/J \leq 10^{-2}$ to form. The relevant energy scale is even smaller for the $m_z = 2/3$ plateau observed in Ref.~\cite{ghosh2025simplexcrystalgroundstate}, as we do not reach temperatures sufficiently low to resolve that plateau, although our ground state calculation does resolve it. We do, however, witness the heat capacity and susceptibility beginning to spike in the vicinity of the $2/3$ plateau at temperatures $T/J \sim 5 \times 10^{-3}$.
We note that the width of the $m_{z} = 1/3$ plateau was observed to be larger than the $m_{z} = 1/2$ plateau in Ref.~\cite{ghosh2025simplexcrystalgroundstate}, but we see the opposite. One would typically expect more commensurate fillings to be more stable to thermal fluctuations, as we do here, and we note that our low temperature energies in the vicinity of the $m_{z} = 1/2$ plateau are lower than those in Ref.~\cite{ghosh2025simplexcrystalgroundstate} and are both converged with bond dimension and affected negligibly by a loop correction (see Appendix, Fig. \ref{Fig:CTMRGComparisons}).

A number of other notable features are manifest in our results here. First, the heat capacity is rapidly increasing with decreasing temperature around $T/J \sim 5 \times 10^{-3}$ for field strengths $h \sim J$ and $h \sim 3J$ and the system is starting to lose most---but not all---of its remaining entropy.
This suggests that further, much more thermally fragile short-range liquid states may be beginning to appear which could be resolved at lower temperatures and finer values of the field. We thus anticipate that the magnetic response of the system could follow a devil's staircase pattern involving a fractal pattern of increasingly narrow magnetic plateaus as has been observed in various frustrated pyrochlore systems~\cite{kondakor2023crystalline, pal2019magnetization}.

At low temperatures, in the vicinity ($h = 2.025J$) of the $m_{z} = 1/2$ and $m_{z} = 1/3$ plateaus, meanwhile, we observe the system switching between these two plateaus, with the heat capacity remaining finite but appearing to sharpen towards a discontinuity with increasing bond dimension (see Fig. \ref{Fig:SpecificMags}A) leading to an asymmetric `lambda' style peak. In this regime, the magnetic susceptibility is also very large. The system effectively appears to be switching between a liquid state on the edge of the $1/2$ plateau to another, less magnetic, liquid state near the $1/3$ plateau. The $T = 0$ edges of these plateaus likely correspond to first- and second-order quantum phase transitions, respectively: the $m_z = 1/2$ edge ($h \sim 2.2J$) appears discontinuous in the magnetization, whereas the $m_z = 1/3$ edge ($h \sim 1.9J$) appears continuous (red dotted line, Fig. \ref{Fig:HeatMaps}). The finite-temperature anomaly we observe may therefore reflect the critical fluctuations of the continuous $1/3$ transition and/or the proximity of a finite-temperature endpoint of the first-order $1/2$ line, which warrants future investigation.

\paragraph*{Conclusion and outlook. ---}
We have mapped out the finite-temperature phase diagram of the spin$-1/2$ Heisenberg antiferromagnet in a magnetic field on the ruby lattice. By leveraging an iTNS ansatz optimized and measured within the belief propagation framework, we have accurately computed thermodynamic quantities of interest, such as the magnetization density, the energy density, the heat capacity, and the entropy across a wide range of temperatures. Such a controlled finite-temperature study gives substantial insights into the formation of the magnetic plateaus and reveals the disordered liquid-like nature of the states within these plateaus. We provide reliable estimates of the energy gap associated with these states as well as strong evidence for their extensive degeneracy.

Our work validates tensor networks plus loop-corrected belief propagation as a powerful framework for studying finite temperature properties of frustrated models in two and three dimensions.
Possible extensions to the present work include identifying perturbations which break the ground state degeneracy in favor of topologically ordered superpositions of crystalline simplex states and moving into three dimensions to study the pyrochlore lattice, where our methodology can be readily utilized.

\subsection*{Acknowledgments}
The authors are grateful for ongoing support by the Flatiron Institute and would like to acknowledge Alexander Wietek, Paul Ebert, and Roderich Moessner for insightful discussions.
The results in this work were produced with the Julia package  
TensorNetworkQuantumSimulator.jl \cite{TensorNetworkQuantumSimulator}, an open source wrapper built on ITensors.jl \cite{fishman2022itensor} for simulating quantum systems with tensor networks of arbitrary topology.
\bibliography{bibliography}

\begin{thebibliography}{42}%
\makeatletter
\providecommand \@ifxundefined [1]{%
 \@ifx{#1\undefined}
}%
\providecommand \@ifnum [1]{%
 \ifnum #1\expandafter \@firstoftwo
 \else \expandafter \@secondoftwo
 \fi
}%
\providecommand \@ifx [1]{%
 \ifx #1\expandafter \@firstoftwo
 \else \expandafter \@secondoftwo
 \fi
}%
\providecommand \natexlab [1]{#1}%
\providecommand \enquote  [1]{``#1''}%
\providecommand \bibnamefont  [1]{#1}%
\providecommand \bibfnamefont [1]{#1}%
\providecommand \citenamefont [1]{#1}%
\providecommand \href@noop [0]{\@secondoftwo}%
\providecommand \href [0]{\begingroup \@sanitize@url \@href}%
\providecommand \@href[1]{\@@startlink{#1}\@@href}%
\providecommand \@@href[1]{\endgroup#1\@@endlink}%
\providecommand \@sanitize@url [0]{\catcode `\\12\catcode `\$12\catcode `\&12\catcode `\#12\catcode `\^12\catcode `\_12\catcode `\%12\relax}%
\providecommand \@@startlink[1]{}%
\providecommand \@@endlink[0]{}%
\providecommand \url  [0]{\begingroup\@sanitize@url \@url }%
\providecommand \@url [1]{\endgroup\@href {#1}{\urlprefix }}%
\providecommand \urlprefix  [0]{URL }%
\providecommand \Eprint [0]{\href }%
\providecommand \doibase [0]{https://doi.org/}%
\providecommand \selectlanguage [0]{\@gobble}%
\providecommand \bibinfo  [0]{\@secondoftwo}%
\providecommand \bibfield  [0]{\@secondoftwo}%
\providecommand \translation [1]{[#1]}%
\providecommand \BibitemOpen [0]{}%
\providecommand \bibitemStop [0]{}%
\providecommand \bibitemNoStop [0]{.\EOS\space}%
\providecommand \EOS [0]{\spacefactor3000\relax}%
\providecommand \BibitemShut  [1]{\csname bibitem#1\endcsname}%
\let\auto@bib@innerbib\@empty
\bibitem [{\citenamefont {Wen}(2004)}]{wen2004quantum}%
  \BibitemOpen
  \bibfield  {author} {\bibinfo {author} {\bibfnamefont {X.-G.}\ \bibnamefont {Wen}},\ }\href {https://doi.org/10.1093/acprof:oso/9780199227259.001.0001} {\emph {\bibinfo {title} {Quantum field theory of many-body systems}}}\ (\bibinfo  {publisher} {Oxford University Press},\ \bibinfo {year} {2004})\BibitemShut {NoStop}%
\bibitem [{\citenamefont {Savary}\ and\ \citenamefont {Balents}(2017)}]{savary2017quantum}%
  \BibitemOpen
  \bibfield  {author} {\bibinfo {author} {\bibfnamefont {L.}~\bibnamefont {Savary}}\ and\ \bibinfo {author} {\bibfnamefont {L.}~\bibnamefont {Balents}},\ }\bibfield  {title} {\bibinfo {title} {Quantum spin liquids: a review},\ }\href {https://doi.org/10.1088/0034-4885/80/1/016502} {\bibfield  {journal} {\bibinfo  {journal} {Reports on Progress in Physics}\ }\textbf {\bibinfo {volume} {80}},\ \bibinfo {pages} {016502} (\bibinfo {year} {2017})}\BibitemShut {NoStop}%
\bibitem [{\citenamefont {Read}\ and\ \citenamefont {Sachdev}(1991)}]{read1991frustr}%
  \BibitemOpen
  \bibfield  {author} {\bibinfo {author} {\bibfnamefont {N.}~\bibnamefont {Read}}\ and\ \bibinfo {author} {\bibfnamefont {S.}~\bibnamefont {Sachdev}},\ }\bibfield  {title} {\bibinfo {title} {Large-$n$ expansion for frustrated quantum antiferromagnets},\ }\href {https://doi.org/10.1103/PhysRevLett.66.1773} {\bibfield  {journal} {\bibinfo  {journal} {Phys. Rev. Lett.}\ }\textbf {\bibinfo {volume} {66}},\ \bibinfo {pages} {1773} (\bibinfo {year} {1991})}\BibitemShut {NoStop}%
\bibitem [{\citenamefont {Sachdev}(1992)}]{sachdev1992antif}%
  \BibitemOpen
  \bibfield  {author} {\bibinfo {author} {\bibfnamefont {S.}~\bibnamefont {Sachdev}},\ }\bibfield  {title} {\bibinfo {title} {{Kagom{\'{e}}}- and triangular-lattice {Heisenberg} antiferromagnets: Ordering from quantum fluctuations and quantum-disordered ground states with unconfined bosonic spinons},\ }\href {https://doi.org/10.1103/PhysRevB.45.12377} {\bibfield  {journal} {\bibinfo  {journal} {Phys. Rev. B}\ }\textbf {\bibinfo {volume} {45}},\ \bibinfo {pages} {12377} (\bibinfo {year} {1992})}\BibitemShut {NoStop}%
\bibitem [{\citenamefont {Anderson}(1987)}]{anderson1987resonating}%
  \BibitemOpen
  \bibfield  {author} {\bibinfo {author} {\bibfnamefont {P.~W.}\ \bibnamefont {Anderson}},\ }\bibfield  {title} {\bibinfo {title} {The resonating valence bond state in {La}$_2${CuO}$_4$ and superconductivity},\ }\href {https://doi.org/10.1126/science.235.4793.1196} {\bibfield  {journal} {\bibinfo  {journal} {Science}\ }\textbf {\bibinfo {volume} {235}},\ \bibinfo {pages} {1196} (\bibinfo {year} {1987})}\BibitemShut {NoStop}%
\bibitem [{\citenamefont {Balents}\ \emph {et~al.}(1998)\citenamefont {Balents}, \citenamefont {Fisher},\ and\ \citenamefont {Nayak}}]{balents1998nodal}%
  \BibitemOpen
  \bibfield  {author} {\bibinfo {author} {\bibfnamefont {L.~S.}\ \bibnamefont {Balents}}, \bibinfo {author} {\bibfnamefont {M.~P.~A.}\ \bibnamefont {Fisher}},\ and\ \bibinfo {author} {\bibfnamefont {C.}~\bibnamefont {Nayak}},\ }\bibfield  {title} {\bibinfo {title} {Nodal liquid theory of the pseudo-gap phase of high-$t_c$ superconductors},\ }\href {https://doi.org/10.1142/S0217979298000570} {\bibfield  {journal} {\bibinfo  {journal} {International Journal of Modern Physics B}\ }\textbf {\bibinfo {volume} {12}},\ \bibinfo {pages} {1033} (\bibinfo {year} {1998})}\BibitemShut {NoStop}%
\bibitem [{\citenamefont {Ghosh}\ and\ \citenamefont {Mila}(2025)}]{ghosh2025simplexcrystalgroundstate}%
  \BibitemOpen
  \bibfield  {author} {\bibinfo {author} {\bibfnamefont {P.}~\bibnamefont {Ghosh}}\ and\ \bibinfo {author} {\bibfnamefont {F.}~\bibnamefont {Mila}},\ }\href {https://arxiv.org/abs/2512.14173} {\bibinfo {title} {Simplex crystal ground state and magnetization plateaus in the spin-$1/2$ {Heisenberg} model on the {Ruby} lattice}} (\bibinfo {year} {2025}),\ \Eprint {https://arxiv.org/abs/2512.14173} {arXiv:2512.14173 [cond-mat.str-el]} \BibitemShut {NoStop}%
\bibitem [{\citenamefont {Maity}\ \emph {et~al.}(2024)\citenamefont {Maity}, \citenamefont {Samajdar},\ and\ \citenamefont {Iqbal}}]{maity2024gappedgaplessquantumspin}%
  \BibitemOpen
  \bibfield  {author} {\bibinfo {author} {\bibfnamefont {A.}~\bibnamefont {Maity}}, \bibinfo {author} {\bibfnamefont {R.}~\bibnamefont {Samajdar}},\ and\ \bibinfo {author} {\bibfnamefont {Y.}~\bibnamefont {Iqbal}},\ }\href {https://arxiv.org/abs/2409.16344} {\bibinfo {title} {Gapped and gapless quantum spin liquids on the {Ruby} lattice}} (\bibinfo {year} {2024}),\ \Eprint {https://arxiv.org/abs/2409.16344} {arXiv:2409.16344 [cond-mat.str-el]} \BibitemShut {NoStop}%
\bibitem [{\citenamefont {Schmoll}\ \emph {et~al.}(2024)\citenamefont {Schmoll}, \citenamefont {Naumann}, \citenamefont {Weerda}, \citenamefont {Eisert},\ and\ \citenamefont {Iqbal}}]{schmoll2024bathing}%
  \BibitemOpen
  \bibfield  {author} {\bibinfo {author} {\bibfnamefont {P.}~\bibnamefont {Schmoll}}, \bibinfo {author} {\bibfnamefont {J.}~\bibnamefont {Naumann}}, \bibinfo {author} {\bibfnamefont {E.~L.}\ \bibnamefont {Weerda}}, \bibinfo {author} {\bibfnamefont {J.}~\bibnamefont {Eisert}},\ and\ \bibinfo {author} {\bibfnamefont {Y.}~\bibnamefont {Iqbal}},\ }\href {https://arxiv.org/abs/2407.07145} {\bibinfo {title} {Bathing in a sea of candidate quantum spin liquids: From the gapless ruby to the gapped maple-leaf lattice}} (\bibinfo {year} {2024}),\ \Eprint {https://arxiv.org/abs/2407.07145} {arXiv:2407.07145 [cond-mat.str-el]} \BibitemShut {NoStop}%
\bibitem [{\citenamefont {Farnell}\ \emph {et~al.}(2011)\citenamefont {Farnell}, \citenamefont {Darradi}, \citenamefont {Schmidt},\ and\ \citenamefont {Richter}}]{farnell2011spin}%
  \BibitemOpen
  \bibfield  {author} {\bibinfo {author} {\bibfnamefont {D.~J.~J.}\ \bibnamefont {Farnell}}, \bibinfo {author} {\bibfnamefont {R.}~\bibnamefont {Darradi}}, \bibinfo {author} {\bibfnamefont {R.}~\bibnamefont {Schmidt}},\ and\ \bibinfo {author} {\bibfnamefont {J.}~\bibnamefont {Richter}},\ }\bibfield  {title} {\bibinfo {title} {Spin-half {Heisenberg} antiferromagnet on two {Archimedean} lattices: From the bounce lattice to the maple-leaf lattice and beyond},\ }\href {https://doi.org/10.1103/PhysRevB.84.104406} {\bibfield  {journal} {\bibinfo  {journal} {Physical Review B}\ }\textbf {\bibinfo {volume} {84}},\ \bibinfo {pages} {104406} (\bibinfo {year} {2011})}\BibitemShut {NoStop}%
\bibitem [{\citenamefont {Sch{\"{a}}fer}\ \emph {et~al.}(2023)\citenamefont {Sch{\"{a}}fer}, \citenamefont {Placke}, \citenamefont {Benton},\ and\ \citenamefont {Moessner}}]{schafer2023abundance}%
  \BibitemOpen
  \bibfield  {author} {\bibinfo {author} {\bibfnamefont {R.}~\bibnamefont {Sch{\"{a}}fer}}, \bibinfo {author} {\bibfnamefont {B.}~\bibnamefont {Placke}}, \bibinfo {author} {\bibfnamefont {O.}~\bibnamefont {Benton}},\ and\ \bibinfo {author} {\bibfnamefont {R.}~\bibnamefont {Moessner}},\ }\bibfield  {title} {\bibinfo {title} {Abundance of hard-hexagon crystals in the quantum pyrochlore antiferromagnet},\ }\href {https://doi.org/10.1103/PhysRevLett.131.096702} {\bibfield  {journal} {\bibinfo  {journal} {Physical Review Letters}\ }\textbf {\bibinfo {volume} {131}},\ \bibinfo {pages} {096702} (\bibinfo {year} {2023})}\BibitemShut {NoStop}%
\bibitem [{\citenamefont {Moessner}\ and\ \citenamefont {Chalker}(1998)}]{moessner98spinliquid}%
  \BibitemOpen
  \bibfield  {author} {\bibinfo {author} {\bibfnamefont {R.}~\bibnamefont {Moessner}}\ and\ \bibinfo {author} {\bibfnamefont {J.~T.}\ \bibnamefont {Chalker}},\ }\bibfield  {title} {\bibinfo {title} {Properties of a classical spin liquid: The {Heisenberg} pyrochlore antiferromagnet},\ }\href {https://doi.org/10.1103/PhysRevLett.80.2929} {\bibfield  {journal} {\bibinfo  {journal} {Phys. Rev. Lett.}\ }\textbf {\bibinfo {volume} {80}},\ \bibinfo {pages} {2929} (\bibinfo {year} {1998})}\BibitemShut {NoStop}%
\bibitem [{\citenamefont {Gardner}\ \emph {et~al.}(2010)\citenamefont {Gardner}, \citenamefont {Gingras},\ and\ \citenamefont {Greedan}}]{gardner2010magnetic}%
  \BibitemOpen
  \bibfield  {author} {\bibinfo {author} {\bibfnamefont {J.~S.}\ \bibnamefont {Gardner}}, \bibinfo {author} {\bibfnamefont {M.~J.~P.}\ \bibnamefont {Gingras}},\ and\ \bibinfo {author} {\bibfnamefont {J.~E.}\ \bibnamefont {Greedan}},\ }\bibfield  {title} {\bibinfo {title} {Magnetic pyrochlore oxides},\ }\href {https://doi.org/10.1103/RevModPhys.82.53} {\bibfield  {journal} {\bibinfo  {journal} {Reviews of Modern Physics}\ }\textbf {\bibinfo {volume} {82}},\ \bibinfo {pages} {53} (\bibinfo {year} {2010})}\BibitemShut {NoStop}%
\bibitem [{\citenamefont {Iqbal}\ \emph {et~al.}(2019)\citenamefont {Iqbal}, \citenamefont {M{\"{u}}ller}, \citenamefont {Ghosh}, \citenamefont {Gingras}, \citenamefont {Jeschke}, \citenamefont {Rachel}, \citenamefont {Reuther},\ and\ \citenamefont {Thomale}}]{iqbal2019quantum}%
  \BibitemOpen
  \bibfield  {author} {\bibinfo {author} {\bibfnamefont {Y.}~\bibnamefont {Iqbal}}, \bibinfo {author} {\bibfnamefont {T.}~\bibnamefont {M{\"{u}}ller}}, \bibinfo {author} {\bibfnamefont {P.~G.}\ \bibnamefont {Ghosh}}, \bibinfo {author} {\bibfnamefont {M.~J.~P.}\ \bibnamefont {Gingras}}, \bibinfo {author} {\bibfnamefont {H.~O.}\ \bibnamefont {Jeschke}}, \bibinfo {author} {\bibfnamefont {S.}~\bibnamefont {Rachel}}, \bibinfo {author} {\bibfnamefont {J.}~\bibnamefont {Reuther}},\ and\ \bibinfo {author} {\bibfnamefont {R.}~\bibnamefont {Thomale}},\ }\bibfield  {title} {\bibinfo {title} {Quantum and classical phases of the pyrochlore {Heisenberg} model with competing interactions},\ }\href {https://doi.org/10.1103/PhysRevX.9.011005} {\bibfield  {journal} {\bibinfo  {journal} {Physical Review X}\ }\textbf {\bibinfo {volume} {9}},\ \bibinfo {pages} {011005} (\bibinfo {year} {2019})}\BibitemShut {NoStop}%
\bibitem [{\citenamefont {Semeghini}\ \emph {et~al.}(2021)\citenamefont {Semeghini}, \citenamefont {Levine}, \citenamefont {Keesling}, \citenamefont {Ebadi}, \citenamefont {Wang}, \citenamefont {Bluvstein}, \citenamefont {Verresen}, \citenamefont {Pichler}, \citenamefont {Kalinowski}, \citenamefont {Samajdar} \emph {et~al.}}]{semeghini2021probing}%
  \BibitemOpen
  \bibfield  {author} {\bibinfo {author} {\bibfnamefont {G.}~\bibnamefont {Semeghini}}, \bibinfo {author} {\bibfnamefont {H.}~\bibnamefont {Levine}}, \bibinfo {author} {\bibfnamefont {A.}~\bibnamefont {Keesling}}, \bibinfo {author} {\bibfnamefont {S.}~\bibnamefont {Ebadi}}, \bibinfo {author} {\bibfnamefont {T.~T.}\ \bibnamefont {Wang}}, \bibinfo {author} {\bibfnamefont {D.}~\bibnamefont {Bluvstein}}, \bibinfo {author} {\bibfnamefont {R.}~\bibnamefont {Verresen}}, \bibinfo {author} {\bibfnamefont {H.}~\bibnamefont {Pichler}}, \bibinfo {author} {\bibfnamefont {M.~w.}\ \bibnamefont {Kalinowski}}, \bibinfo {author} {\bibfnamefont {R.}~\bibnamefont {Samajdar}}, \emph {et~al.},\ }\bibfield  {title} {\bibinfo {title} {Probing topological spin liquids on a programmable quantum simulator},\ }\href {https://doi.org/10.1126/science.abi8794} {\bibfield  {journal} {\bibinfo  {journal} {Science}\ }\textbf {\bibinfo {volume} {374}},\ \bibinfo {pages} {1242} (\bibinfo {year} {2021})}\BibitemShut {NoStop}%
\bibitem [{\citenamefont {Giudici}\ \emph {et~al.}(2022)\citenamefont {Giudici}, \citenamefont {Lukin},\ and\ \citenamefont {Pichler}}]{giudici2022dynamical}%
  \BibitemOpen
  \bibfield  {author} {\bibinfo {author} {\bibfnamefont {G.}~\bibnamefont {Giudici}}, \bibinfo {author} {\bibfnamefont {M.~D.}\ \bibnamefont {Lukin}},\ and\ \bibinfo {author} {\bibfnamefont {H.}~\bibnamefont {Pichler}},\ }\bibfield  {title} {\bibinfo {title} {Dynamical preparation of quantum spin liquids in {Rydberg} atom arrays},\ }\href {https://doi.org/10.1103/PhysRevLett.129.090401} {\bibfield  {journal} {\bibinfo  {journal} {Physical Review Letters}\ }\textbf {\bibinfo {volume} {129}},\ \bibinfo {pages} {090401} (\bibinfo {year} {2022})}\BibitemShut {NoStop}%
\bibitem [{\citenamefont {Tarabunga}\ \emph {et~al.}(2022)\citenamefont {Tarabunga}, \citenamefont {Surace}, \citenamefont {Andreoni}, \citenamefont {Angelone},\ and\ \citenamefont {Dalmonte}}]{tarabunga2022gauge}%
  \BibitemOpen
  \bibfield  {author} {\bibinfo {author} {\bibfnamefont {P.~S.}\ \bibnamefont {Tarabunga}}, \bibinfo {author} {\bibfnamefont {F.~M.}\ \bibnamefont {Surace}}, \bibinfo {author} {\bibfnamefont {R.}~\bibnamefont {Andreoni}}, \bibinfo {author} {\bibfnamefont {A.}~\bibnamefont {Angelone}},\ and\ \bibinfo {author} {\bibfnamefont {M.}~\bibnamefont {Dalmonte}},\ }\bibfield  {title} {\bibinfo {title} {Gauge-theoretic origin of {Rydberg} quantum spin liquids},\ }\href {https://doi.org/10.1103/PhysRevLett.129.195301} {\bibfield  {journal} {\bibinfo  {journal} {Physical Review Letters}\ }\textbf {\bibinfo {volume} {129}},\ \bibinfo {pages} {195301} (\bibinfo {year} {2022})}\BibitemShut {NoStop}%
\bibitem [{\citenamefont {Verresen}\ \emph {et~al.}(2021)\citenamefont {Verresen}, \citenamefont {Lukin},\ and\ \citenamefont {Vishwanath}}]{verresen2021prediction}%
  \BibitemOpen
  \bibfield  {author} {\bibinfo {author} {\bibfnamefont {R.}~\bibnamefont {Verresen}}, \bibinfo {author} {\bibfnamefont {M.~D.}\ \bibnamefont {Lukin}},\ and\ \bibinfo {author} {\bibfnamefont {A.}~\bibnamefont {Vishwanath}},\ }\bibfield  {title} {\bibinfo {title} {Prediction of toric code topological order from {Rydberg} blockade},\ }\href {https://doi.org/10.1103/PhysRevX.11.031005} {\bibfield  {journal} {\bibinfo  {journal} {Physical Review X}\ }\textbf {\bibinfo {volume} {11}},\ \bibinfo {pages} {031005} (\bibinfo {year} {2021})}\BibitemShut {NoStop}%
\bibitem [{\citenamefont {Jeschke}\ \emph {et~al.}(2026)\citenamefont {Jeschke}, \citenamefont {Guterding},\ and\ \citenamefont {Ghosh}}]{jeschke2026}%
  \BibitemOpen
  \bibfield  {author} {\bibinfo {author} {\bibfnamefont {H.~O.}\ \bibnamefont {Jeschke}}, \bibinfo {author} {\bibfnamefont {D.}~\bibnamefont {Guterding}},\ and\ \bibinfo {author} {\bibfnamefont {P.}~\bibnamefont {Ghosh}},\ }\href {https://arxiv.org/abs/2605.28821} {\bibinfo {title} {Realization of the ruby lattice antiferromagnet in layered transition-metal fluorides}} (\bibinfo {year} {2026}),\ \Eprint {https://arxiv.org/abs/2605.28821} {arXiv:2605.28821 [cond-mat.str-el]} \BibitemShut {NoStop}%
\bibitem [{\citenamefont {Jiang}\ \emph {et~al.}(2008)\citenamefont {Jiang}, \citenamefont {Weng},\ and\ \citenamefont {Xiang}}]{jiang2008accurate}%
  \BibitemOpen
  \bibfield  {author} {\bibinfo {author} {\bibfnamefont {H.-C.}\ \bibnamefont {Jiang}}, \bibinfo {author} {\bibfnamefont {Z.-Y.}\ \bibnamefont {Weng}},\ and\ \bibinfo {author} {\bibfnamefont {T.}~\bibnamefont {Xiang}},\ }\bibfield  {title} {\bibinfo {title} {Accurate determination of tensor network state of quantum lattice models in two dimensions},\ }\href {https://doi.org/10.1103/PhysRevLett.101.090603} {\bibfield  {journal} {\bibinfo  {journal} {Physical Review Letters}\ }\textbf {\bibinfo {volume} {101}},\ \bibinfo {pages} {090603} (\bibinfo {year} {2008})}\BibitemShut {NoStop}%
\bibitem [{\citenamefont {Nishino}\ and\ \citenamefont {Okunishi}(1996)}]{nishino1996corner}%
  \BibitemOpen
  \bibfield  {author} {\bibinfo {author} {\bibfnamefont {T.}~\bibnamefont {Nishino}}\ and\ \bibinfo {author} {\bibfnamefont {K.}~\bibnamefont {Okunishi}},\ }\bibfield  {title} {\bibinfo {title} {Corner transfer matrix renormalization group method},\ }\href {https://doi.org/10.1143/JPSJ.65.891} {\bibfield  {journal} {\bibinfo  {journal} {Journal of the Physical Society of Japan}\ }\textbf {\bibinfo {volume} {65}},\ \bibinfo {pages} {891} (\bibinfo {year} {1996})}\BibitemShut {NoStop}%
\bibitem [{\citenamefont {Or{\'{u}}s}\ and\ \citenamefont {Vidal}(2009)}]{orus2009simulation}%
  \BibitemOpen
  \bibfield  {author} {\bibinfo {author} {\bibfnamefont {R.}~\bibnamefont {Or{\'{u}}s}}\ and\ \bibinfo {author} {\bibfnamefont {G.}~\bibnamefont {Vidal}},\ }\bibfield  {title} {\bibinfo {title} {Simulation of two-dimensional quantum systems on an infinite lattice revisited: Corner transfer matrix for tensor contraction},\ }\href {https://doi.org/10.1103/PhysRevB.80.094403} {\bibfield  {journal} {\bibinfo  {journal} {Physical Review B}\ }\textbf {\bibinfo {volume} {80}},\ \bibinfo {pages} {094403} (\bibinfo {year} {2009})}\BibitemShut {NoStop}%
\bibitem [{\citenamefont {Yang}\ and\ \citenamefont {Corboz}(2026)}]{Yang_2026}%
  \BibitemOpen
  \bibfield  {author} {\bibinfo {author} {\bibfnamefont {Q.}~\bibnamefont {Yang}}\ and\ \bibinfo {author} {\bibfnamefont {P.}~\bibnamefont {Corboz}},\ }\bibfield  {title} {\bibinfo {title} {Efficient i{PEPS} simulation on the honeycomb lattice via {QR}-based corner transfer matrix renormalization group},\ }\bibfield  {journal} {\bibinfo  {journal} {Physical Review B}\ }\textbf {\bibinfo {volume} {113}},\ \href {https://doi.org/10.1103/9gmp-byx8} {10.1103/9gmp-byx8} (\bibinfo {year} {2026})\BibitemShut {NoStop}%
\bibitem [{\citenamefont {Kasteleyn}(1961)}]{Kastelyn1961}%
  \BibitemOpen
  \bibfield  {author} {\bibinfo {author} {\bibfnamefont {P.~W.}\ \bibnamefont {Kasteleyn}},\ }\bibfield  {title} {\bibinfo {title} {The statistics of dimers on a lattice: I. the number of dimer arrangements on a quadratic lattice},\ }\href {https://doi.org/10.1016/0031-8914(61)90063-5} {\bibfield  {journal} {\bibinfo  {journal} {Physica}\ }\textbf {\bibinfo {volume} {27}},\ \bibinfo {pages} {1209} (\bibinfo {year} {1961})}\BibitemShut {NoStop}%
\bibitem [{\citenamefont {Temperley}\ and\ \citenamefont {Fisher}(1961)}]{Temperley1961}%
  \BibitemOpen
  \bibfield  {author} {\bibinfo {author} {\bibfnamefont {H.~N.~V.}\ \bibnamefont {Temperley}}\ and\ \bibinfo {author} {\bibfnamefont {M.~E.}\ \bibnamefont {Fisher}},\ }\bibfield  {title} {\bibinfo {title} {Dimer problem in statistical mechanics---an exact result},\ }\href {https://doi.org/10.1080/14786436108243366} {\bibfield  {journal} {\bibinfo  {journal} {The Philosophical Magazine: A Journal of Theoretical Experimental and Applied Physics}\ }\textbf {\bibinfo {volume} {6}},\ \bibinfo {pages} {1061} (\bibinfo {year} {1961})}\BibitemShut {NoStop}%
\bibitem [{\citenamefont {Alkabetz}\ and\ \citenamefont {Arad}(2021)}]{alkabetz2021tensor}%
  \BibitemOpen
  \bibfield  {author} {\bibinfo {author} {\bibfnamefont {R.}~\bibnamefont {Alkabetz}}\ and\ \bibinfo {author} {\bibfnamefont {I.}~\bibnamefont {Arad}},\ }\bibfield  {title} {\bibinfo {title} {Tensor networks contraction and the belief propagation algorithm},\ }\href {https://doi.org/10.1103/PhysRevResearch.3.023073} {\bibfield  {journal} {\bibinfo  {journal} {Physical Review Research}\ }\textbf {\bibinfo {volume} {3}},\ \bibinfo {pages} {023073} (\bibinfo {year} {2021})}\BibitemShut {NoStop}%
\bibitem [{\citenamefont {Tindall}\ and\ \citenamefont {Fishman}(2023)}]{tindall2023gauging}%
  \BibitemOpen
  \bibfield  {author} {\bibinfo {author} {\bibfnamefont {J.}~\bibnamefont {Tindall}}\ and\ \bibinfo {author} {\bibfnamefont {M.}~\bibnamefont {Fishman}},\ }\bibfield  {title} {\bibinfo {title} {Gauging tensor networks with belief propagation},\ }\href {https://doi.org/10.21468/SciPostPhys.15.6.222} {\bibfield  {journal} {\bibinfo  {journal} {SciPost Phys.}\ }\textbf {\bibinfo {volume} {15}},\ \bibinfo {pages} {222} (\bibinfo {year} {2023})}\BibitemShut {NoStop}%
\bibitem [{\citenamefont {Evenbly}\ \emph {et~al.}(2026)\citenamefont {Evenbly}, \citenamefont {Pancotti}, \citenamefont {Milsted}, \citenamefont {Gray},\ and\ \citenamefont {Chan}}]{evenbly2026loop}%
  \BibitemOpen
  \bibfield  {author} {\bibinfo {author} {\bibfnamefont {G.}~\bibnamefont {Evenbly}}, \bibinfo {author} {\bibfnamefont {N.}~\bibnamefont {Pancotti}}, \bibinfo {author} {\bibfnamefont {A.}~\bibnamefont {Milsted}}, \bibinfo {author} {\bibfnamefont {J.}~\bibnamefont {Gray}},\ and\ \bibinfo {author} {\bibfnamefont {G.~K.-L.}\ \bibnamefont {Chan}},\ }\bibfield  {title} {\bibinfo {title} {Loop series expansions for tensor networks},\ }\href {https://doi.org/10.1103/vqks-cr6x} {\bibfield  {journal} {\bibinfo  {journal} {Physical Review Research}\ }\textbf {\bibinfo {volume} {8}},\ \bibinfo {pages} {013245} (\bibinfo {year} {2026})}\BibitemShut {NoStop}%
\bibitem [{\citenamefont {Evenbly}\ \emph {et~al.}(2025)\citenamefont {Evenbly}, \citenamefont {Gray},\ and\ \citenamefont {Chan}}]{evenbly2025partitioned}%
  \BibitemOpen
  \bibfield  {author} {\bibinfo {author} {\bibfnamefont {G.}~\bibnamefont {Evenbly}}, \bibinfo {author} {\bibfnamefont {J.}~\bibnamefont {Gray}},\ and\ \bibinfo {author} {\bibfnamefont {G.~K.-L.}\ \bibnamefont {Chan}},\ }\href {https://arxiv.org/abs/2512.10910} {\bibinfo {title} {Partitioned expansions for approximate tensor network contractions}} (\bibinfo {year} {2025}),\ \Eprint {https://arxiv.org/abs/2512.10910} {arXiv:2512.10910 [quant-ph]} \BibitemShut {NoStop}%
\bibitem [{\citenamefont {Guo}\ \emph {et~al.}(2023)\citenamefont {Guo}, \citenamefont {Poletti},\ and\ \citenamefont {Arad}}]{guo2023block}%
  \BibitemOpen
  \bibfield  {author} {\bibinfo {author} {\bibfnamefont {C.}~\bibnamefont {Guo}}, \bibinfo {author} {\bibfnamefont {D.}~\bibnamefont {Poletti}},\ and\ \bibinfo {author} {\bibfnamefont {I.}~\bibnamefont {Arad}},\ }\bibfield  {title} {\bibinfo {title} {Block belief propagation algorithm for two-dimensional tensor networks},\ }\href {https://doi.org/10.1103/PhysRevB.108.125111} {\bibfield  {journal} {\bibinfo  {journal} {Physical Review B}\ }\textbf {\bibinfo {volume} {108}},\ \bibinfo {pages} {125111} (\bibinfo {year} {2023})}\BibitemShut {NoStop}%
\bibitem [{\citenamefont {Midha}\ and\ \citenamefont {Zhang}(2025)}]{midha2025beyond}%
  \BibitemOpen
  \bibfield  {author} {\bibinfo {author} {\bibfnamefont {S.}~\bibnamefont {Midha}}\ and\ \bibinfo {author} {\bibfnamefont {Y.~F.}\ \bibnamefont {Zhang}},\ }\href {https://arxiv.org/abs/2510.02290} {\bibinfo {title} {Beyond {Belief} {Propagation}: Cluster-corrected tensor network contraction with exponential convergence}} (\bibinfo {year} {2025}),\ \Eprint {https://arxiv.org/abs/2510.02290} {arXiv:2510.02290 [quant-ph]} \BibitemShut {NoStop}%
\bibitem [{\citenamefont {Midha}\ \emph {et~al.}(2026)\citenamefont {Midha}, \citenamefont {Sommers}, \citenamefont {Tindall},\ and\ \citenamefont {Abanin}}]{midha2026belief}%
  \BibitemOpen
  \bibfield  {author} {\bibinfo {author} {\bibfnamefont {S.}~\bibnamefont {Midha}}, \bibinfo {author} {\bibfnamefont {G.~M.}\ \bibnamefont {Sommers}}, \bibinfo {author} {\bibfnamefont {J.}~\bibnamefont {Tindall}},\ and\ \bibinfo {author} {\bibfnamefont {D.~A.}\ \bibnamefont {Abanin}},\ }\href {https://arxiv.org/abs/2604.03228} {\bibinfo {title} {{Belief} {Propagation} and tensor network expansions for many-body quantum systems: Rigorous results and fundamental limits}} (\bibinfo {year} {2026}),\ \Eprint {https://arxiv.org/abs/2604.03228} {arXiv:2604.03228 [quant-ph]} \BibitemShut {NoStop}%
\bibitem [{\citenamefont {Tindall}\ \emph {et~al.}(2026)\citenamefont {Tindall}, \citenamefont {Mello}, \citenamefont {Fishman}, \citenamefont {Stoudenmire},\ and\ \citenamefont {Sels}}]{tindall2026}%
  \BibitemOpen
  \bibfield  {author} {\bibinfo {author} {\bibfnamefont {J.}~\bibnamefont {Tindall}}, \bibinfo {author} {\bibfnamefont {A.~F.}\ \bibnamefont {Mello}}, \bibinfo {author} {\bibfnamefont {M.}~\bibnamefont {Fishman}}, \bibinfo {author} {\bibfnamefont {E.~M.}\ \bibnamefont {Stoudenmire}},\ and\ \bibinfo {author} {\bibfnamefont {D.}~\bibnamefont {Sels}},\ }\bibfield  {title} {\bibinfo {title} {Dynamics of disordered quantum systems with two- and three-dimensional tensor networks},\ }\href {https://doi.org/10.1126/science.adx2728} {\bibfield  {journal} {\bibinfo  {journal} {Science}\ }\textbf {\bibinfo {volume} {392}},\ \bibinfo {pages} {868} (\bibinfo {year} {2026})},\ \Eprint {https://arxiv.org/abs/https://www.science.org/doi/pdf/10.1126/science.adx2728} {https://www.science.org/doi/pdf/10.1126/science.adx2728} \BibitemShut {NoStop}%
\bibitem [{\citenamefont {Sch{\"{a}}fer}\ \emph {et~al.}(2026)\citenamefont {Sch{\"{a}}fer}, \citenamefont {Ebert}, \citenamefont {Hassan}, \citenamefont {Reuther}, \citenamefont {Luitz},\ and\ \citenamefont {Wietek}}]{Sch_fer_2026}%
  \BibitemOpen
  \bibfield  {author} {\bibinfo {author} {\bibfnamefont {R.}~\bibnamefont {Sch{\"{a}}fer}}, \bibinfo {author} {\bibfnamefont {P.~L.}\ \bibnamefont {Ebert}}, \bibinfo {author} {\bibfnamefont {N.}~\bibnamefont {Hassan}}, \bibinfo {author} {\bibfnamefont {J.}~\bibnamefont {Reuther}}, \bibinfo {author} {\bibfnamefont {D.~J.}\ \bibnamefont {Luitz}},\ and\ \bibinfo {author} {\bibfnamefont {A.}~\bibnamefont {Wietek}},\ }\bibfield  {title} {\bibinfo {title} {Thermodynamics of the {Heisenberg} antiferromagnet on the maple-leaf lattice},\ }\bibfield  {journal} {\bibinfo  {journal} {Zeitschrift f{\"{u}}r Naturforschung A}\ }\href {https://doi.org/10.1515/zna-2025-0382} {10.1515/zna-2025-0382} (\bibinfo {year} {2026})\BibitemShut {NoStop}%
\bibitem [{\citenamefont {Ebert}\ \emph {et~al.}(2026)\citenamefont {Ebert}, \citenamefont {Iqbal},\ and\ \citenamefont {Wietek}}]{ebert2026}%
  \BibitemOpen
  \bibfield  {author} {\bibinfo {author} {\bibfnamefont {P.~L.}\ \bibnamefont {Ebert}}, \bibinfo {author} {\bibfnamefont {Y.}~\bibnamefont {Iqbal}},\ and\ \bibinfo {author} {\bibfnamefont {A.}~\bibnamefont {Wietek}},\ }\href {https://arxiv.org/abs/2605.12592} {\bibinfo {title} {Incommensurate spin-density waves in a frustrated maple-leaf lattice ferromagnet}} (\bibinfo {year} {2026}),\ \Eprint {https://arxiv.org/abs/2605.12592} {arXiv:2605.12592 [cond-mat.str-el]} \BibitemShut {NoStop}%
\bibitem [{Note1()}]{Note1}%
  \BibitemOpen
  \bibinfo {note} {All derivatives in this work are computed using a second-order finite-difference scheme.}\BibitemShut {Stop}%
\bibitem [{\citenamefont {Pierre}\ \emph {et~al.}(2024)\citenamefont {Pierre}, \citenamefont {Bernu},\ and\ \citenamefont {Messio}}]{pierre2024high}%
  \BibitemOpen
  \bibfield  {author} {\bibinfo {author} {\bibfnamefont {L.}~\bibnamefont {Pierre}}, \bibinfo {author} {\bibfnamefont {B.}~\bibnamefont {Bernu}},\ and\ \bibinfo {author} {\bibfnamefont {L.}~\bibnamefont {Messio}},\ }\bibfield  {title} {\bibinfo {title} {High temperature series expansions of $s=1/2$ {Heisenberg} spin models: algorithm to include the magnetic field with optimized complexity},\ }\href {https://doi.org/10.21468/SciPostPhys.17.4.105} {\bibfield  {journal} {\bibinfo  {journal} {SciPost Physics}\ }\textbf {\bibinfo {volume} {17}},\ \bibinfo {pages} {105} (\bibinfo {year} {2024})}\BibitemShut {NoStop}%
\bibitem [{\citenamefont {Kond{\'{a}}kor}\ and\ \citenamefont {Penc}(2023)}]{kondakor2023crystalline}%
  \BibitemOpen
  \bibfield  {author} {\bibinfo {author} {\bibfnamefont {M.}~\bibnamefont {Kond{\'{a}}kor}}\ and\ \bibinfo {author} {\bibfnamefont {K.}~\bibnamefont {Penc}},\ }\bibfield  {title} {\bibinfo {title} {Crystalline phases and devil's staircase in qubit spin ice},\ }\href {https://doi.org/10.1103/PhysRevResearch.5.043172} {\bibfield  {journal} {\bibinfo  {journal} {Physical Review Research}\ }\textbf {\bibinfo {volume} {5}},\ \bibinfo {pages} {043172} (\bibinfo {year} {2023})}\BibitemShut {NoStop}%
\bibitem [{\citenamefont {Pal}\ and\ \citenamefont {Lal}(2019)}]{pal2019magnetization}%
  \BibitemOpen
  \bibfield  {author} {\bibinfo {author} {\bibfnamefont {S.}~\bibnamefont {Pal}}\ and\ \bibinfo {author} {\bibfnamefont {S.}~\bibnamefont {Lal}},\ }\bibfield  {title} {\bibinfo {title} {Magnetization plateaus of the quantum pyrochlore {Heisenberg} antiferromagnet},\ }\href {https://doi.org/10.1103/PhysRevB.100.104421} {\bibfield  {journal} {\bibinfo  {journal} {Physical Review B}\ }\textbf {\bibinfo {volume} {100}},\ \bibinfo {pages} {104421} (\bibinfo {year} {2019})}\BibitemShut {NoStop}%
\bibitem [{Ten(2026)}]{TensorNetworkQuantumSimulator}%
  \BibitemOpen
  \href@noop {} {\bibinfo {title} {{{TensorNetworkQuantumSimulator.jl}}}},\ \bibinfo {howpublished} {\url{https://github.com/JoeyT1994/TensorNetworkQuantumSimulator}} (\bibinfo {year} {2026})\BibitemShut {NoStop}%
\bibitem [{\citenamefont {Fishman}\ \emph {et~al.}(2022)\citenamefont {Fishman}, \citenamefont {White},\ and\ \citenamefont {Stoudenmire}}]{fishman2022itensor}%
  \BibitemOpen
  \bibfield  {author} {\bibinfo {author} {\bibfnamefont {M.}~\bibnamefont {Fishman}}, \bibinfo {author} {\bibfnamefont {S.}~\bibnamefont {White}},\ and\ \bibinfo {author} {\bibfnamefont {E.}~\bibnamefont {Stoudenmire}},\ }\bibfield  {title} {\bibinfo {title} {The {ITensor} software library for tensor network calculations},\ }\href {https://doi.org/10.21468/SciPostPhysCodeb.4-r0.3} {\bibfield  {journal} {\bibinfo  {journal} {SciPost Physics Codebases}\ ,\ \bibinfo {pages} {004}} (\bibinfo {year} {2022})}\BibitemShut {NoStop}%
\bibitem [{\citenamefont {Cao}\ and\ \citenamefont {Vontobel}(2017)}]{Cao_2017}%
  \BibitemOpen
  \bibfield  {author} {\bibinfo {author} {\bibfnamefont {M.~X.}\ \bibnamefont {Cao}}\ and\ \bibinfo {author} {\bibfnamefont {P.~O.}\ \bibnamefont {Vontobel}},\ }\bibfield  {title} {\bibinfo {title} {Double-edge factor graphs: Definition, properties, and examples},\ }in\ \href {https://doi.org/10.1109/itw.2017.8277985} {\emph {\bibinfo {booktitle} {2017 IEEE Information Theory Workshop (ITW)}}}\ (\bibinfo  {publisher} {IEEE},\ \bibinfo {year} {2017})\ pp.\ \bibinfo {pages} {136--140}\BibitemShut {NoStop}%
\end{thebibliography}%

\begin{widetext}
\newpage 
\section*{Appendix}

\subsection*{Belief Propagation for iTNS at finite temperature}
We here provide an extended explanation of the use of the belief propagation (BP) algorithm on the finite temperature iTNS depicted in Fig.~\ref{Fig:h0_panels} of the main text. Let $\rho^{\frac{1}{2}}(\beta)$ be the iTNS representation of the thermal density matrix of the system. It consists of a unit cell of two local tensors (one for each sublattice) $\rho_{a}^{\frac{1}{2}}(\beta)$ and $\rho_{b}^{\frac{1}{2}}(\beta)$, each with six spin degrees of freedom (three bra and three ket for the involved spins) and three virtual degrees of freedom connecting them to their neighboring tensors in the other sublattice. Such a unit cell can be understood as being tiled in an infinite pattern to form the full iTNS, with vertices  $\rho_{v}^{\frac{1}{2}}(\beta)$.

We run the belief propagation algorithm on the `norm' network $Z= {\rm Tr}\left(\rho(\beta)\right)={\rm Tr}\left(\rho^{\frac{1}{2}}(\beta)(\rho^{\frac{1}{2}}(\beta))^{T}\right)$ formed from contracting two copies of $\rho^{\frac{1}{2}}(\beta)$ over all physical indices. 
The local tensors of this norm network are $Z_v = \rho_{v}^{\frac{1}{2}}(\beta) \cdot \left(\rho_{v}^{\frac{1}{2}}(\beta) \right)^{\dagger}$, where the $\cdot$ implies tensor contraction over the six common indices. Since we are working with a $2$-site unit cell, $v \in \{a,b\}$.
We emphasize the choice to represent the square root of $\rho(\beta)$ and contract  $Z={\rm Tr}\left(\rho^{\frac{1}{2}}(\beta)(\rho^{\frac{1}{2}}(\beta))^{T}\right)$ to optimize local tensors and make measurements is very deliberate. This is because the local tensors of the norm network $Z_{v}$ are positive semi-definite (PSD) when interpreted as a map from the virtual indices of the two copies of $\rho_{v}^{\frac{1}{2}}(\beta)$. The PSD property is generally important for numerical stability when approximately contracting tensor networks \cite{Cao_2017}, including with methods such as BP.

To run the BP algorithm on the norm network iTNS, message tensors $M_{v \rightarrow v'}$ are introduced between the vertices of the norm network such that their indices match the two virtual indices of dimension $\chi$. These message tensors are updated according to the self-consistent rule
\begin{equation}\label{Eq:message_up}
    M_{v \rightarrow v'} =\left( \prod_{v''\in ({\rm neighbors}(v)/v')}M_{v'' \rightarrow v}\right)Z_v
\end{equation}
on the lattice which is based on making the Bethe-free energy---a tree-like approximation to $Z$ ---stationary. This update rule is illustrated in Fig. \ref{Fig:LCdiagrams}B.

As we have a two site unit cell, there are only six unique message tensors that need to be specified, three leaving $Z_{a}$ towards each of its $Z_{b}$ neighbors and three leaving $Z_{b}$ towards each of its $Z_{a}$ neighbors. Note we are not assuming $C_{3v}$ symmetry in our tensors and thus we need to assign a message uniquely to each direction leaving a tensor.

For numerical stability and later convenience in computing corrections to BP, we impose the normalization condition $M_{v \to v'} \cdot M_{v'\to v}=1$ after each update. Upon convergence, the messages incident to any contiguous region of the iTNS form a rank-1 approximation to its contraction around that region, directly in the thermodynamic limit.
We use this approximation to condition the truncation down to a bond dimension $\chi$ following the application of imaginary-time gates which lower the temperature of our iTNS. We emphasize that the truncation procedure becomes exact in the limit $\chi \rightarrow \infty$ (i.e.\ the approximation is only used if truncation is performed) and thus our gate application becomes increasingly unbiased by the BP approximation with increased bond dimension. In that limit, any remaining BP approximation is in the measurement of properties of the state.

Once the iTNS at a given temperature and bond dimension has been realized, the finite temperature expectation value of an observable $O = \prod_{v \in V}O_{v}$ with support on some set of sites $V$ can be readily computed under the BP approximation using the converged messages from Eq.~\eqref{Eq:message_up} as
\begin{align}
    \langle O(\beta) \rangle = \lim_{n \rightarrow \infty}\frac{1}{n}\frac{{\rm Tr}\left(\rho(\beta) O\right)}{{\rm Tr}\left(\rho(\beta)\right)} \approx \frac{\prod_{ v \in V}\left( \left(\rho_{v}^{\frac{1}{2}}(\beta) O_v \rho_{v}^{\frac{1}{2}}(\beta) \right)\prod_{ e \in \partial V}M_{e}\right)}{\prod_{ v \in V}\left( \left(\rho_{v}^{\frac{1}{2}}(\beta) \rho_{v}^{\frac{1}{2}}(\beta) \right)\prod_{e \in \partial V}M_{e}\right)}.
    \label{eq:ObsAppendix}
\end{align}
where $\partial V$ is the set of edges incident from the rest of network to the region $V$. If the support of $V$ is larger than our unit cell, we simply have to stitch together copies of the unit cell and messages to span the support and perform the relevant contraction.

With the converged messages, the BP approximation to the free energy, i.e.\ minus the log of the partition function $Z (\beta )={\rm Tr} \left( \exp \left(-\beta H \right)\right) $ can also immediately be read off
\begin{align}
    F_{\rm BP}(\beta)  = -\sum_{v}\log \left(Z_{v}\prod_{ v' \in {\rm neighbors}(v)}M_{v'\rightarrow v}\right)
\end{align}
where we have assumed normalization $M_{v \to v'} \cdot M_{v'\to v}=1$ of messages. As we only have two unique vertices in our unit cell, the free energy density directly follows as

\begin{equation}
    f_{\rm BP}(\beta)  = -\frac{1}{6}\left(\log \left(Z_{a}\prod_{ v' \in {\rm neighbors}(a)}M_{v'\rightarrow a}\right) + \log \left(Z_{b}\prod_{ v' \in {\rm neighbors}(b)}M_{v'\rightarrow b}\right)\right) \approx -\lim_{n \rightarrow \infty}\frac{1}{n}\log ({\rm Tr}(\rho(\beta)))
\end{equation}
with the factor of $\frac{1}{6}$ reflecting the fact we have six spins in our unit cell (three in $a$, three in $b$). This is illustrated in Fig. \ref{Fig:LCdiagrams}.

Therefore, as an alternative to Eq. (\ref{eq:ObsAppendix}) expectation values of thermodynamic observables such as the energy and the magnetization can also be retrieved through simple thermodynamic equalities involving the free energy density $f(\beta) = -\frac{1}{n}\log \left( Z(\beta)\right)$ as
\begin{align}
    e(\beta) = \frac{\partial f(\beta)}{\partial \beta} \ \ \rm{and}\ \ m_z(\beta) =-\frac{2}{\beta}\frac{\partial f(\beta)}{\partial h}.
\end{align}
We always find these to be in agreement with Eq. (\ref{eq:ObsAppendix}).

This latter approach is particularly amenable to the calculation of loop corrections to BP. Loop-corrected expectation values of the observables in the thermodynamic limit can indeed be directly obtained from the loop-corrected value of the free energy density.  At first order, the correction corresponds to contracting a single hexagonal plaquette of alternating $A$ and $B$ tensors together with the incoming fixed-point BP messages attached to each virtual leg. A visual representation of $f(\beta)$ along with its BP value and corrections is displayed in Fig.~\ref{Fig:LCdiagrams} and our loop-corrected observables in this work correspond to taking appropriate derivatives of $f_{\rm BP}(\beta) - \frac{1}{2}W_{6}$ \cite{evenbly2026loop}.

\begin{figure}[t!]
\centering
\includegraphics[width=\textwidth]{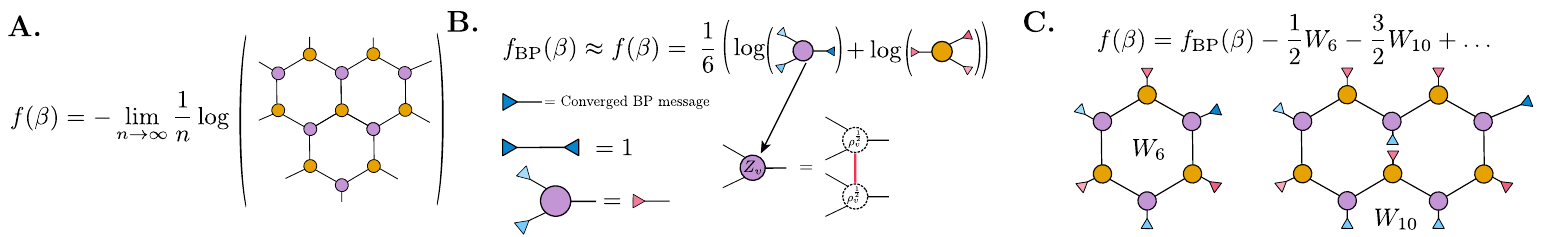}
\caption{\textbf{A)} Tensor diagram for the free energy density in the thermodynamic limit. \textbf{B)} BP approximation of the partition function $Z(\beta)$. The triangles represent converged BP messages, while the solid and dashed tensors are $Z_v$ and $\rho^{\frac{1}{2}}_v$, respectively. The contracted red index in $\rho_v$ accounts for the contraction over the six physical indices on each vertex $v$. \textbf{C)} First and second order correction to the BP value for $f(\beta)$.}
\label{Fig:LCdiagrams}
\end{figure}

\subsection*{Comparison with CTMRG results}
In Fig.~\ref{Fig:CTMRGComparisons} we compare our $T=0.005J$ results obtained with $\chi=96$ with the CTMRG ones in Ref.~\cite{ghosh2025simplexcrystalgroundstate} obtained with $D=12$. In the left panel, we show the percentage relative error $\displaystyle \frac{e_{\mathrm{CTMRG}}-e_{T=0.005}}{e_{\mathrm{CTMRG}}}$ for the energy density $e$. Notice that our energy is higher (within $0.04 \%$) with respect to the CTMRG one across the $m_z =0$ plateau, while being lower in the proximity of the $m_z=1/2$ plateau and for most values of $h$.
In the right panel, we also show the absolute error $m_{\mathrm{CTMRG}}-m_{T=0.005}$ for the magnetization density $m_z$. 
\begin{figure}[h!]
\centering
\includegraphics[width=\textwidth]{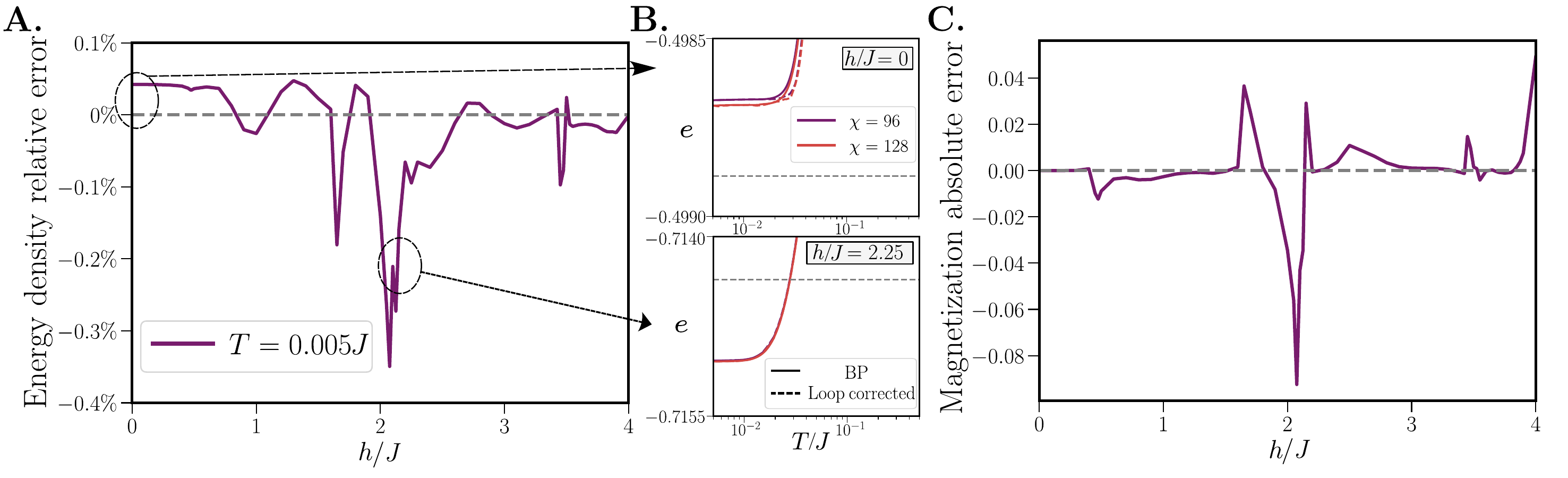}
\caption{\textbf{A)} Percentage relative error in the energy density at $T=0.005J$ measured through BP ($\chi=96$) with respect to the CTMRG ground state results. \textbf{B)} BP and loop corrected energy density $e$ as a function of temperature for $h/J=0, 2.25$ along with the $T=0$ CTMRG estimate. \textbf{C)} absolute error in the magnetization density at $T=0.005J$ measured through BP ($\chi=96$) with respect to the CTMRG ground state results from Ref. \cite{ghosh2025simplexcrystalgroundstate}. A negative value implies the finite temperature energy we obtained via BP is lower than the CTMRG obtained results in Ref. \cite{ghosh2025simplexcrystalgroundstate} and vice versa.}
\label{Fig:CTMRGComparisons}
\end{figure}

\subsection*{Gap estimate for the $m_z = 1/3$ plateau}
The estimate of the gap for the $m_z = 1/3$ plateau is  challenging due to the fact that it forms at temperatures around one order of magnitude lower than the $0$ and $1/2$ plateaus. In Fig.~\ref{Fig:OneThirdGap}, we show the BP results for the specific heat and the entropy as a function of temperature for $h = 1.83J$, which sits within the plateau. Specifically, the inset on the left panel displays the estimate of the gap, which is approximately two and seven times smaller compared with the $1/2$ and $0$ gaps, respectively. 
\begin{figure}[h!]
\centering
\includegraphics[width=0.9\textwidth]{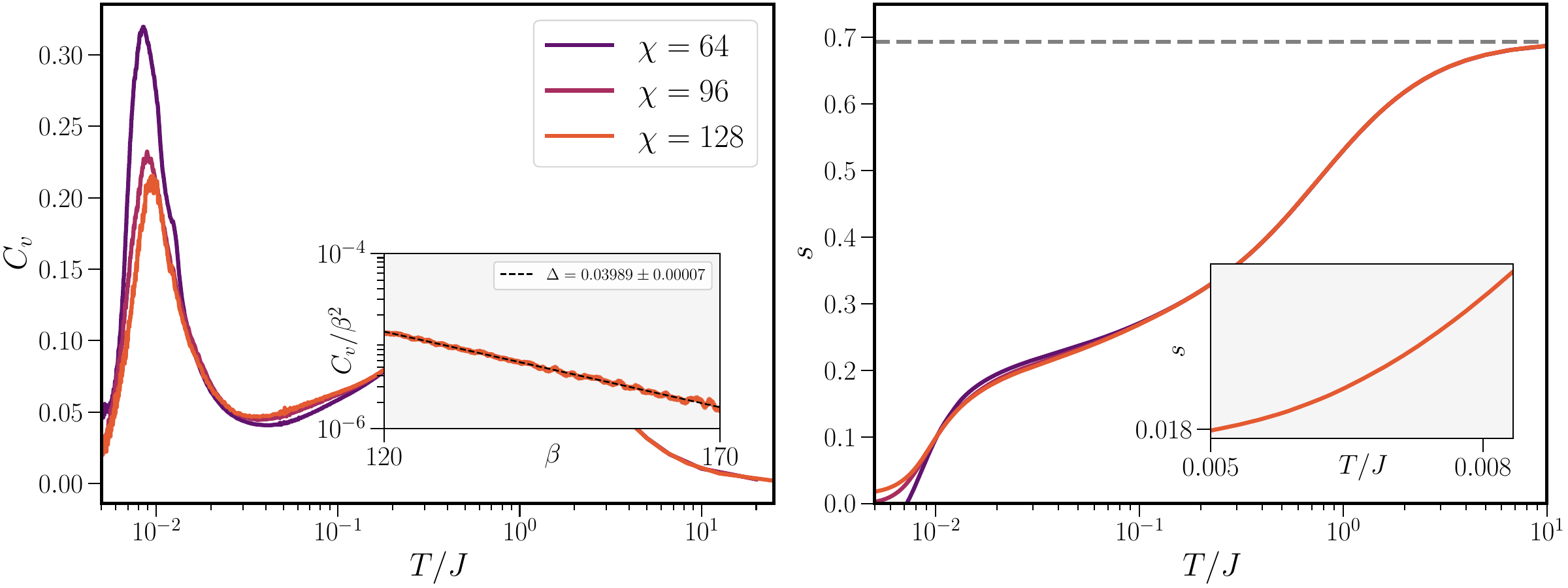}
\caption{Left: specific heat as a function of temperature for $h/J=1.83$ and $\chi =64,96$. Inset: gap estimate. Right: entropy as a function of temperature for $h/J=1.83$ and $\chi=64,96$. Inset: residual entropy.}
\label{Fig:OneThirdGap}
\end{figure}

\subsection*{Correlation length and entanglement}
The accuracy of the BP approximation on the honeycomb geometry relies on the suppression of correlations around the hexagonal loops of the coarse-grained structure.
The simplex crystal states that make up the low energy eigenstates of the model exhibit very weak loop correlations 
due to the fact each hexagon can only contain, at most, three strongly paired spin simplices and the remaining pairs must be weakly correlated.
\begin{figure}[h!]
\centering
\includegraphics[width=0.8\textwidth]{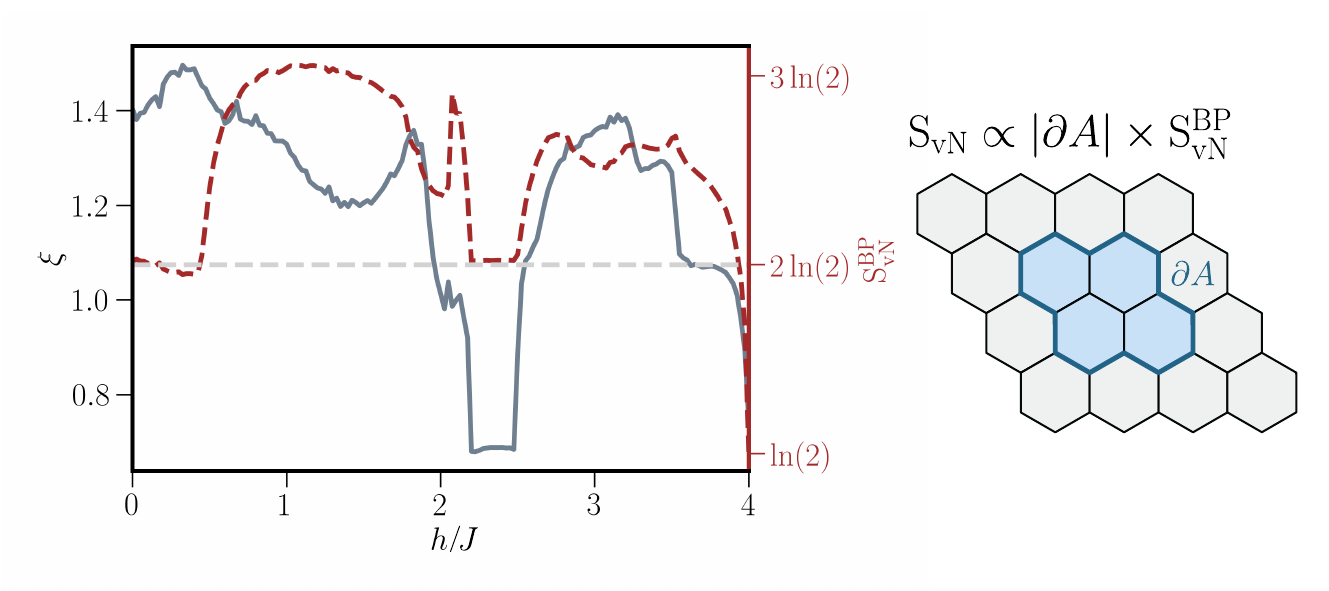}
\caption{Primary axis: correlation length of the thermal state (see Eq. (\ref{eq:AppendixCorrLength}) at $T=5\times 10^{-3}J$ computed from the BP environments. Secondary axis: BP-computed area law coefficient for the operator entanglement entropy for the density matrix $\rho(T)$ at $T=5\times 10^{-3}J$ obtained from BP environments---see Eqs. \ref{Eq:AppendixBPEEV1} and \ref{Eq:AppendixBPEEV2}.}

\label{Fig:corr_vne}
\end{figure}
To make this claim explicit, in Fig. \ref{Fig:corr_vne} we show our calculation of the correlation length
\begin{equation}
    \xi = \frac{1}{\log(\vert \lambda_{1} \vert/\vert\lambda_{2}\vert)}
    \label{eq:AppendixCorrLength}
\end{equation}
from the ordered eigenvalues $\lambda$ of the  $\chi^{2} \times \chi^{2}$ BP transfer matrix for the norm of the thermal iTNS $\rho^{\frac{1}{2}}(\beta)$ across the full range of $h$ at our lowest reached temperature $5 \times 10^{-3}J$. We observe that $\xi$ exhibits signatures of the magnetization plateaus, peaking at their edges and staying constant within. However, it remains notably small throughout, being always on the order of $a$, where $a$ is the inter-simplex spacing (i.e. the length of one bond of the hexagonal lattice). Consequently, the local quantum many-body physics of the model can be mostly understood as a Husimi cactus of triangles of frustrated spins, justifying the BP approach. At low temperatures, however, global properties such as the residual entropy are seen to change with a loop correction---reflecting the fact that the hexagonal lattice imposes loop-specific constraints on the states within the manifold.

We also plot the von-Neumann operator entanglement entropy $\mathrm{S}^{\mathrm{BP}}_{\mathrm{vN}}$ of the density matrix as measured through the BP-approximated environments, and displayed on the secondary $y-$axis in Fig.~\ref{Fig:corr_vne}.
More specifically, for a given edge $e$ of the hexagonal lattice we can compute 
\begin{equation}
    \mathrm{S}^{\mathrm{BP}}_{e} 
    = - \sum_{i = 1}^{\chi^{2}}\lambda_i \log \lambda_i
    \label{Eq:AppendixBPEEV1}
\end{equation}
where $\lbrace \lambda_i \rbrace$ are the eigenvalues of the $\chi^{2} \times \chi^{2}$ matrix $m^{\frac{1}{2}}_{{\rm reverse} (e)} \cdot m_{e} \cdot m^{\frac{1}{2}}_{{\rm reverse} (e)}$ with the dot denoting matrix multiplication (the BP messages have two indices and thus can be interpreted as matrices). For a contiguous region $A$ of the system $S$ enclosed by a border $\partial A$ which is defined by a set of edges then a BP prediction for the vN entanglement entropy of that region is
\begin{equation}
    S^{BP}_{\rm A} =  \sum_{e \in \partial A}S^{BP}_{e} \approx S^{\rm vN}_{A},
        \label{Eq:AppendixBPEEV2}
\end{equation}
where $S^{\rm vN}_{A} =- \sum_{\lambda}\lambda \log(\lambda)$ is the true operator entanglement entropy with $\lambda$ the eigenvalues of the reduced density matrix ${\rm Tr}_{S \setminus A}\left(\rho(\beta)\right)$.
For our infinite setup, and given the translational invariance of the density matrix we find $S^{\rm BP}_{e} = S^{\rm BP}$, i.e.\ it is identical $\forall e$ and so BP predicts $S^{BP}_{A} = S^{\rm BP} \vert \partial A \vert$ and thus gives (given the small correlation length with respect to the size of a hexagon) an accurate approximation coefficient on the area law of operator entanglement entropy. The area law coefficient is shown diagrammatically in the Figure~\ref{Fig:corr_vne}, given a region $A$ (in blue) whose border we denote by $\partial A$. 
As we are working with a mixed state this quantity is operator-based and, interestingly, we see an area law scaling coefficient of almost precisely $\ln(4)$ within the $0$ and $1/2$ plateaus.

\end{widetext}
\end{document}